\documentclass[fleqn,usenatbib]{gji}

\usepackage{newtxtext,newtxmath}
\usepackage[T1]{fontenc}
\usepackage{ae,aecompl}

\usepackage{amsmath}
\usepackage{epic}
\usepackage{graphicx}
\usepackage{hyperref}
\usepackage{multirow}
\usepackage{natbib}
\usepackage{overpic}
\usepackage{color}
\AtBeginShipout{%
 \ifnum\value{page}>1 %
 \typeout{* Additional boxing of page `\thepage'}%
 \setbox\AtBeginShipoutBox=\hbox{\copy\AtBeginShipoutBox}%
 \fi
}
\title[Separation of horizontal GPS station velocities]{Separation of tectonic and local components of horizontal GPS station velocities: a case study for glacial isostatic adjustment in East Antarctica}
\author[Turner et al.]{Ross J. Turner$^{1\thanks{turner.rj@icloud.com}}$, Anya M. Reading$^{1,2}$ and Matt A. King$^{3}$\\
  $^1$ Physics, Private Bag 37, University of Tasmania, Hobart, TAS 7001, Australia \\ $^2$ Institute for Marine and Antarctic Studies, University of Tasmania, Private Bag 129, Hobart, TAS 7001, Australia \\ $^3$ Geography and Spatial Sciences, Private Bag 38, University of Tasmania, Hobart, TAS 7001, Australia
  }

\date{Accepted 2020 May 27. Received 2020 May 27; in original form 2019 December 17}
\pubyear{2020}

\begin{document}

\label{firstpage}

\maketitle

\begin{abstract}
Accurate measurement of the local component of geodetic motion at GPS stations presents a challenge due to the need to separate this signal from the tectonic plate rotation.  A pressing example is the observation of glacial isostatic adjustment (GIA) which constrains the Earth's response to ice unloading, and hence, contributions of ice-covered regions such as Antarctica to global sea level rise following ice mass loss. While both vertical and horizontal motions are of interest in general, we focus on horizontal GPS velocities which typically contain a large component of plate rotation and a smaller {local} component {primarily} relating to GIA.  Incomplete separation of these components introduces significant bias into estimates of GIA motion vectors.
We present the results of a series of tests based on the motions of GPS stations from East Antarctica: 1) signal separation for sets of synthetic data that replicate the geometric character of non-separable, and separable, GIA-like horizontal velocities; and 2) signal separation for real GPS station data with an appraisal of uncertainties.  For both synthetic and real motions, we compare results where the stations are unweighted, and where each station is {areal-}weighted using a metric representing the inverse of the spatial density of neighbouring stations.  %The minimum scale length of assumed geometric variability for the GIA-like signal is controllable in the tests. 
From the synthetic tests, we show that a GIA-like signal is recoverable from the plate rotation signal providing it has geometric variability across East Antarctica. We also show that {areal-}weighting has a very significant effect on the ability to recover a GIA-like signal with geometric variability, and hence on separating the plate rotation and {local} components.  For the real data, assuming a rigid Antarctic plate, fitted plate rotation parameters compare well with other studies in the literature.  %The weighted signal separation test on real data gives a more plausible result with the ability to quantify uncertainty. 
We find that 25 out of 36 GPS stations examined in East Antarctica have {non-zero} {local} horizontal velocities, at the 2$\sigma$ level, after signal separation. We make the code for weighted signal separation available to assist in the consistent appraisal of separated signals, and the comparison of likely uncertainty bounds, for future studies.

\end{abstract}

\begin{keywords}
Glaciology; Plate motions; Antarctica; Satellite geodesy.

\end{keywords}

\section{Introduction}

Measured horizontal GPS velocities comprise a tectonic plate rotation component and {a local} component. At intraplate station sites, the local component is typically much smaller than the plate rotation. In ice-covered or recently deglaciated regions, the local component of Earth deformation is dominated by the signal arising from glacial isostatic adjustment (GIA). Separation of this signal is vital, as an example, to successfully interpret GRACE satellite data in order to estimate the ice mass contribution towards sea level rise from the given study region \citep{King+2012}. The errors potentially introduced due to the incomplete separation of plate rotation and local signals \citep{Klemann+2008, King+2016} are significant and {complicate} the use of horizontal GPS velocities. Local horizontal motions, importantly, are sensitive to lateral variations in Earth rheology and promise considerable advances in understanding interactions and feedbacks between the solid Earth and ice sheets \citep{vanderWal+2015} if they can be successfully isolated. These horizontal site velocities, such as those due to GIA, are typically separated from the plate rotation by making use of a stable reference location unaffected by glacial rebound or seismic events {\citep{Johansson+2002, Sella+2007, Altamimi+2012}}. This technique {has been} used successfully in Greenland and the European Alps \citep[e.g.][]{Barletta+2006}, though is not possible for the entirely ice-covered Antarctic Plate {\citep[e.g.][]{Argus+2014} or even the large North American plate as there is significant far-field motions \citep{Kreemer+2018}}.

The ITRF2014 plate rotation model \citep{Altamimi+2017} defines an absolute plate motion for 11 major tectonic plates, including Antarctica, from 318 sites believed to be under very limited influence from local effects such as plate boundary and other deformation zones. The bedrock topography of the Antarctic continent is divided into two distinct regions by the Transantarctic Mountains \citep[see e.g. Bedmap2 data of][]{Fretwell+2013}; to the east the bedrock extends up to 3000 metres above mean sea level whereas the West Antarctic Rift System immediately to the west is submerged below sea level. The sites used in the ITRF2014 model to define the motion of the Antarctic Plate are all located in East Antarctica; West Antarctic sites were excluded due to their proximity to a plate boundary or being located in a region of extreme GIA such as the land area adjacent to the Larsen Ice Shelf \citep[e.g.][]{Nield+2014, Barletta+2018}. Moreover, the Antarctic Peninsula is known to recede from East Antarctica at a rate of 1-$2\rm\, mm/yr$ towards South Georgia \citep{Siddoway+2008}, widening the West Antarctic Rift which formed from 115-$100\rm\, Ma$. 

Although the motions are much smaller, East Antarctic GPS station sites are also subject to local motions attributable to GIA. The GIA vector field should theoretically be recoverable from the combined velocity field if the length-scale of local motions is less than {that of} the plate motion%, i.e. the curl of the velocity field explains at most a small fraction of the GIA velocities ($\rm{curl}\, \boldsymbol{F} = \nabla \times \boldsymbol{F}$; an operator that measures the rotational component of a vector field $\boldsymbol{F}$)
. \citet{Argus+2014} attempt to constrain both the GIA signal and plate motion in Antarctica as part of a postglacial rebound study by including only the relatively stable sites in East Antarctica and the Kerguelen Islands. The magnitudes of their recovered GIA velocities are consistent with physically-based models which consider the surface load history during the last glacial cycle and variations in Earth rheology \citep{Whitehouse+2012}. However, the success of the separation of the plate rotation and GIA signals in studies of the Antarctic continent remains in doubt \citep{King+2016}, with the expected errors encountered using the techniques employed in the literature being largely unquantified.

Localised deformation following the 1998 $M_{\rm w} = 8.1$ Antarctic intraplate earthquake has been shown to persist along parts of the East Antarctic coastline some 600 km removed from the epicentre \citep{King+2016b}. In particular, the horizontal velocities near Dumont d'Urville at present (2009-2017) are up to 2 mm/yr greater in the eastward direction than prior to the 1998 earthquake \citep{King+2016b}. \citet{DeMets+2016} have similarly suggested that Casey Station is affected by this earthquake and the 2004 $M_{\rm z} = 9.3$ Sumatra earthquake, though this might be partly explained by elastic deformation due to melting of the Totten Glacier \citep{Roberts+2018}. Modelling of viscoelastic {postseismic} deformation predicts that much of Antarctica may still be deforming from this and other large seismic events along the Macquarie Ridge \citep{Nettles+1999, Watson+2010}. This mode of Antarctic deformation needs to be considered (due to its large magnitude relative to typical GIA signals) in the selection of sites used in geodetic estimates of plate motion and GIA.  

In this contribution, we investigate the separation of plate rotation and local components of horizontal GPS station velocities using synthetic and real data. Our approach draws on the likelihood that the local signals are uncorrelated over longer length-scales and provides a means of determining whether the recovered local signal at each station is significant. As our case example, we use East Antarctic GPS stations which are sparsely distributed, with some denser clusters. For East Antarctica, recovered horizontal signals are appraised as to their ability to define GIA and hence, constrain ice unloading. Our study also informs the deployment of new GPS stations for maximum benefit.

\section{Data and Methods}

\subsection{GPS dataset}

The real, observed, dataset comprises an updated compilation of GPS station motions from permanent installations and {long-term} field campaigns. These data are obtained from public datasets, such as those held by UNAVCO and Geoscience Australia, or provided directly to M.A.K. Full information is provided in Supplement A. {W}e considered GPS data over the period 2009-2013 inclusive, chosen to maximise the common data period following a period of rapid site deployments across Antarctica%, and used the subsequent GPS time series to compute horizontal velocities
. The median data span is 4.0 years. GPS coordinate time series were produced using standard approaches as described by, for instance, \citet{Wolstencroft+2015}. We adopted a Precise Point Positioning (PPP) analysis approach within GIPSY v6.3 software using NASA JPL clocks and orbit products. We transformed the daily coordinate time series into the International GNSS Service realisation of the ITRF2008 \citep{Altamimi+2011}. Site {horizontal} velocities were then derived using CATS v3.1.2 software \citep{Williams+2008}, estimating offsets, annual and semi-annual period terms {(Supplement B)}, and linear trend along with parameters for a white noise plus power-law noise model. {Long-wavelength common mode errors (CME) may lead to moderate changes in our measured horizontal velocities, especially for time-series shorter than five years \citep{Serpelloni+2013}. {However, we accept this more minor source of error as the focus of this work is to investigate techniques to separate the tectonic and local velocity components, rather than to present the final word on GIA velocities in Antarctica.}

{West Antarctica and the Antarctic Peninsula are subject to neotectonic processes and internal deformation in the present epoch \citep{Jordan+2020}. We therefore take a conservative approach and restrict our analysis to stations in East Antarctic.}
Specifically, we include 36 GPS stations at latitudes below $60^\circ$S and at all longitudes except those covering West Antarctica and the Antarctica Peninsula between $40$-$160^\circ$W \citep[cf. Figure 2 of][]{Harley+2013}. 
These 36 East Antarctic stations are widely distributed around the coastline of the continental land mass but also include some regional campaigns (e.g. Transantarctic Mountains) with more closely spaced stations. The dataset of the velocities at these stations (Supplement A) includes the station latitude, longitude, elevation, northward and eastward velocity components, with 1$\sigma$ measurement errors provided for the velocities. The GPS velocity measurements at Dumont d'Urville are supplemented by time series data from the International DORIS Service \citep[\url{http://ids-doris.org/}; processed in][]{King+2016b} to constrain the signal before the the onset of postseismic deformation from the nearby $M_{\rm w} = 8.1$ earthquake in 1998.

\subsection{Signal separation method}

\label{sec:meth}

The movement of the {tectonic} plate is known to be the main contributor to the horizontal component of the GPS station velocities measured in East Antarctica. We assume that the plate is effectively rigid and hence {its motion} can be described as the curl of the {measured} velocity field about {some} Euler pole. GIA contributes a further horizontal motion vector that is typically no larger than 5-10\% of the magnitude of the site motion due to plate rotation \citep{Peltier+2015}. The GIA velocity field is assumed to vary {on a smaller spatial scale} than the plate rotation across an ice covered continental area. {For example, there are likely to be} opposing velocity vectors either side of a centre of ice mass reduction, resulting in site velocities with a divergent character.  

In this section, we summarise the method used to derive horizontal velocities at each GPS station due to tectonic plate motion, and develop a technique to find the {set of Euler pole parameters (describing this motion)} which best explain the {rotational component} of the measured horizontal velocity vector field. This measured horizontal velocity contains both plate rotation and GIA components, and potentially other factors including postseismic deformation{; we describe the calculation of a local velocity vector (i.e. signal with plate rotation subtracted) which we assume largely measures the GIA component of the signal}. 
{We choose to use a Monte Carlo simulation to solve for the Euler pole parameters, and corresponding tectonic and local component of the motion at each station, and provide a full treatment of the uncertainties (i.e. variances, covariances{, and equivalent non-Gaussian measures of spread}). %\textbf{The general technique applied is as follows:}
%\begin{enumerate}
%\item in each Monte Carlo realisation, analytically calculate:
%	\begin{itemize}
%	\item best fit Euler pole parameters for this realisation,
%	\item corresponding tectonic plate motion at each station,
%	\item local component of the motion at each station.
%	\end{itemize}
%\item combining the results from many Monte Carlo realisations, numerically calculate the mean, variance and covariance for:
%	\begin{itemize}
%	\item Euler pole parameters for the Antarctic Plate,
%	\item local component of the motion at each station.
%	\end{itemize}
%\end{enumerate}
%\textbf{The Monte Carlo based least-squares optmisation is explained in more detail in Section \ref{sec:Regression analysis}, whilst a weighting based on the spatial density of stations is described in Secion \ref{sec:Areal weighting}.}

\subsubsection{Unweighted regression}
\label{sec:Regression analysis}

{The total velocity vector for a station located at some latitude and longitude $(\theta, \phi)$ on the surface of a tectonic plate is typically split into a vertical and two horizontal components. In this work, we only consider motion along the Earth's surface of the form $\boldsymbol{v} = v_\theta \boldsymbol{e_\theta} + v_\phi \boldsymbol{e_\phi}$ for northward and eastward components $v_\theta$ and $v_\phi$ respectively; $\boldsymbol{e_\theta}$ and $\boldsymbol{e_\phi}$ are the corresponding northward and eastward unit vectors. The horizontal velocities at a given station can be related to the plate motion using two parameterisations: (1) angular rate of rotation $\omega$ about an arbitrarily located Euler pole $(\theta_\epsilon, \phi_\epsilon)$, where $\theta_\epsilon$ and $\phi_\epsilon$ are the latitude and longitude of the pole \citep[e.g.][]{Goudarzi+2014}; and (2) angular rate of rotation about each of the three cartesian coordinate axes $x$, $y$ and $z$ (i.e. $\omega_{\rm x}$, $\omega_{\rm y}$, $\omega_{\rm z}$) \citep[e.g.][]{McKenzie+1974}. We present both of these parameterisations for the tectonic plate motion in this work for ease of comparision with other authors.}
 %The key equations used in this work to convert between angular rotation about an Euler pole and horizontal velocities on the plate surface are summarised in Supplement B.

%The curl of the velocity field is assumed to be due solely to the plate rotation, as outlined previously. 
The {modelled tectonic plate motion} which best explains the {rotational component of the measured} horizontal velocities is fitted {numerically} using least-squares {optimisation for an arbitrary initial guess} (referred to hereafter as `{unweighted} regression' in text). The {least-squares optimisation} minimises the difference between the measured horizontal velocity vector and the modelled {northward and eastward} component velocities derived {either} for each trial set of rotation rate and Euler pole coordinates (i.e. $\theta_\epsilon, \phi_\epsilon, \omega$) {or for each trial set of rotation rates about the three cartesian coordinate axes (i.e. $\omega_{\rm x}$, $\omega_{\rm y}$, $\omega_{\rm z}$)}. {Note that although it is simple to convert between these parameterisations using standard trigonometric expressions, the conversion between their error distributions is a highly non-linear process.}

{Monte Carlo simulations are used to numerically propagate the m}easurement uncertainties in the horizontal velocities through to the fitted rotation rate and the location of the Euler pole, {or alternatively  to the fitted rotation rates about the cartesian coordinate axes}. The velocities at each GPS station are randomly selected from {the error distribution for the measurement} at that station. The best fitting plate rotation for this random variation of the horizontal velocities across the GPS sites in the Antarctic continent is found using the {unweighted} regression. This process is repeated 5000 times, returning {non-}Gaussian distributions in the best fits for the rotation rate, and Euler pole latitude and longitude {(or alternatively rotation rates about the cartesian coordinate axes)}. The median of these distributions are taken as the best estimate for each parameter, and the 16th and 84th percentiles are assumed for the uncertainties {(in the special case of a Gaussian distribution these percentiles correspond to the $\pm 1\sigma$ confidence level)}. Confidence ellipses describing the covariances similarly use these percentiles to represent a $1\sigma$ confidence level. {Variance-covariance matrices are also calculated for completeness using standard techniques under the assumption of Gaussian uncertainties.}

The relative importance of each GPS station in the {optimisation} may be modified by weighting the least-squares residuals for each station by a coefficient (referred to hereafter as `weighted regression', see Section \ref{sec:Areal weighting}); i.e. ${w_{\rm i}}^2 (\boldsymbol{v}_{\rm i} - \hat{\boldsymbol{v}}_{\rm i})^2$, where $\boldsymbol{v}_{\rm i}$ is the measured horizontal velocity vector at the $i$'th station, $\hat{\boldsymbol{v}}_{\rm i}$ is the analytically predicted velocity, and $w_{\rm i} = 1/\sigma_{\rm i}$ is the weighting applied to this station. This technique to weight the regression is not used to propagate the measurement uncertainties since the Monte Carlo simulation provides a full treatment {of both non-Gaussian uncertainties and the corresponding covariance}. 

{The local component of the signal at each station is taken as the difference between the measured horizontal velocity vector and the plate rotation expected from the fitted {set of Euler pole parameters (i.e. $\theta_\epsilon, \phi_\epsilon, \omega$ or $\omega_{\rm x}$, $\omega_{\rm y}$, $\omega_{\rm z}$)} for each Monte Carlo realisation{; this leads to a distribution of 5000 realisations of the local component of the velocity at each station.} The medians of the resulting distributions in the northward and eastward horizontal velocities are taken as the best estimates, and the 16th and 84th percentiles are again assumed for the uncertainties.} {Confidence ellipses describing the covariances based on the full numerical soultion and variance-covariance matrices assuming Gaussian uncertainties are also calculated.}

\subsubsection{Areal-weighted regression}
\label{sec:Areal weighting}

East Antarctic GPS sites are clustered along the Transantarctic Mountains on the western edge of the Ross Sea (25 of 36 stations); moreover half of these stations are close to either Mount Erebus or McMurdo Station. Horizontal motion due to GIA at stations in close proximity (likely to be on the same side of a centre of ice mass loss) is expected to lead to horizontal velocities in roughly the same direction. This large fraction of stations with likely similar GIA velocity vectors will bias the fitted rotation about the Euler pole; in order to minimise residuals across these 25 stations the fitting process will explain correlated GIA velocities in this region as plate rotation. The fit of the Euler pole parameters must be made using an approximately uniform distribution of GPS sites to prevent bias from proximate stations with correlated GIA.

\begin{figure}
\centerline{\includegraphics[width=0.75\columnwidth,trim={80 50 60 50},clip]{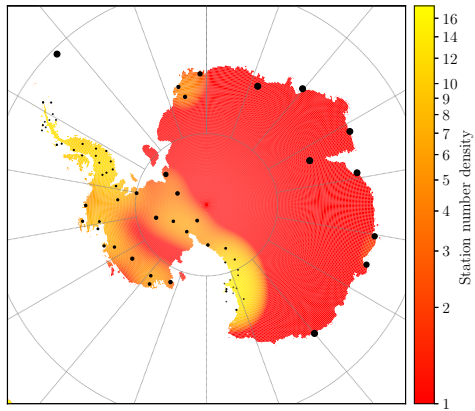}}
\caption{The local number density of GPS sites across Antarctica measured by smoothing the discrete station distribution with a Gaussian filter of FWMH (full-width at half-maximum) length-scale 5 degrees. The point size corresponds to the relative weighting given to each station whilst the colour map shows the local number density of stations across the continent (units of inverse length-scale squared; i.e. $[560\rm\, km]^{-2}$).}
\label{fig:Number_densities}
\end{figure}

The density of GPS stations across the continent is quantified by defining the length-scale over which the GIA signal is expected to be correlated, then smoothing over the discrete station distribution using an appropriately scaled Gaussian filter. This length-scale is taken as the half-power point of the smoothing Gaussian filter; i.e. a GPS site at this separation contributes the equivalent of half a station to the local number density. The least-squares regression is weighted by the inverse of the local number density (i.e. the {effective} area covered by each station) so that GPS sites in close proximity have less influence on the fit than their more isolated counterparts. 

The typical correlation length-scale in the GIA velocity field is investigated using physically informed models, following the method of \citet{King+2016}. This GIA modelling considers both vertical and lateral variations in Earth rheology \citep{vanderWal+2015}, and uses the W12 Antarctic deglaciation model \citep{Whitehouse+2012} to define the surface load history during the last glacial cycle. The lithosphere response is modelled as purely elastic deformation above a depth of 120 km \citep{Latychev+2005} and viscoelastic deformation below this layer; mantle temperature is a major factor in the diffusion and dislocation creep at this depth \citep{vanderWal+2010}, and is modelled using velocity perturbations from different global seismic models. However, the average grain size, water content, and melt content are also important factors \citep{Hirth+2013}. \citet{King+2016} assume{d} no melt content, an average grain size of either 4 or 10 mm, and water content between either somewhat wet (1000 ppm H2O) or a fully dry state. The direction of GIA velocities predicted at each station is highly dependent on model assumptions \citep{King+2016}. However, these models consistently show no correlation between the GIA velocities at Transantarctic Mountains stations and those along the East Antarctic coast, whilst the set of models in which the rheological properties vary only in the vertical direction predict the velocity field becomes anticorrelated over the 1000 km length of the mountains (from RAMG to TNB1). 

{Based on these findings, we assume a correlation length-scale of 5 degrees (corresponding to a diameter of 560km). The local number density of GPS stations across the Antarctic continent is shown in Figure \ref{fig:Number_densities} for our assumed correlation length-scale. We note that the results in this contribution are unaffected by moderate changes in the size of the correlation length-scale, as discussed in Section \ref{sec:antarctic}.}

\section{Separation of velocity components in simulated data}
\label{sec:simulated}

Velocity fields combining simulated plate rotation for the Antarctic continent with different, simulated `GIA' vector fields are used in this section to test the separation of the two signals. Horizontal motions at the locations of 36 GPS stations in East Antarctica (a description of station selection is given in Section \ref{sec:antarctic}) are constructed assuming the plate rotation found by \citet[][$\omega = 0.224^\circ/\rm Ma$, $\theta_\epsilon = 58.69^\circ \rm\, N$, $\phi_\epsilon = -128.29^\circ \rm\, E$]{Wei-Ping+2009}.

The simulated horizontal `GIA' vector fields are assigned appropriate magnitudes guided by the work of \citet{Peltier+2015}. These `GIA' vectors are thus scaled to a median magnitude of 1 mm/yr and are directed either: (1) following the rotation about the Euler pole (i.e. curl); (2) normal to the rotation about the Euler pole; (3) radially outwards from Dome A on the Antarctic Plateau (i.e. divergence; see Figure \ref{fig:Simulated_GIA}); or (4) random spherical harmonic functions for the two velocity components each with randomly generated order in latitude for $v_\theta$ and in longitude for $v_\phi$. 

The Laplace (real valued) spherical harmonics are simulated with random orientation to the Earth's geographic poles. The degree of the harmonics is assumed to be a random integer between 18 and 36 for both the latitudinal and longitudinal velocity components, corresponding to a correlation length-scale of 5 to 10 degrees. Each spherical harmonic order is also sampled with equal weighting to include a mix of zonal, sectoral and tesseral variability. These assumptions ensure vectors are correlated on the order of a few degrees (the characteristic length-scale assumed for the GIA field) whilst maintaining a quasi-random field across the broader continent. The velocity field is simulated for 5000 random variations in the degree, order and orientation of the spherical harmonics.

\subsection{Unweighted regression analysis}

The {unweighted} regression is applied to the four simulated datasets of horizontal velocities (constructed as above) for the locations of 36 East Antarctic GPS stations. The recovered `GIA' fields, separated from the plate motion, are shown in Figure \ref{fig:Simulated_GIA}. If the added `GIA' fields are entirely parallel, Figure \ref{fig:Simulated_GIA} (a), or entirely normal (b) to the rotation about the Euler pole then it is not possible to separate this signal from the plate rotation (any inferred plate rotation will be shifted accordingly). {The `GIA' component at a given station can always be interpreted as rotation about some Euler pole, but it is only in the special case of highly correlated signals (as above) that these Euler poles converge to a common location, exactly as expected for a rigid plate}. For the field aligned with the divergence from Dome A (c), the recovered `GIA' velocities capture approximately half of the simulated magnitudes, however, the residuals of median magnitude $\sim$0.57 mm/yr (see Table \ref{tab:gia_velocity}) all show a direction approximately parallel to the 20th meridian east. This suggests the regression has incorrectly assumed some of the `GIA' velocity field results from the rotation of the Antarctic continent about an Euler pole. This arises from the concentration of GPS locations along the Transantarctic Mountains in this simulation, which present a very asymmetric distribution about the geometric centre of the divergence (i.e. Dome A). The poorly recovered (too small) `GIA' velocities at any cluster of stations leads to the residuals at these sites pointing directly towards centre of divergence.

\begin{figure*}
\centerline{\includegraphics[width=1.25\columnwidth,trim={105 105 85 100},clip]{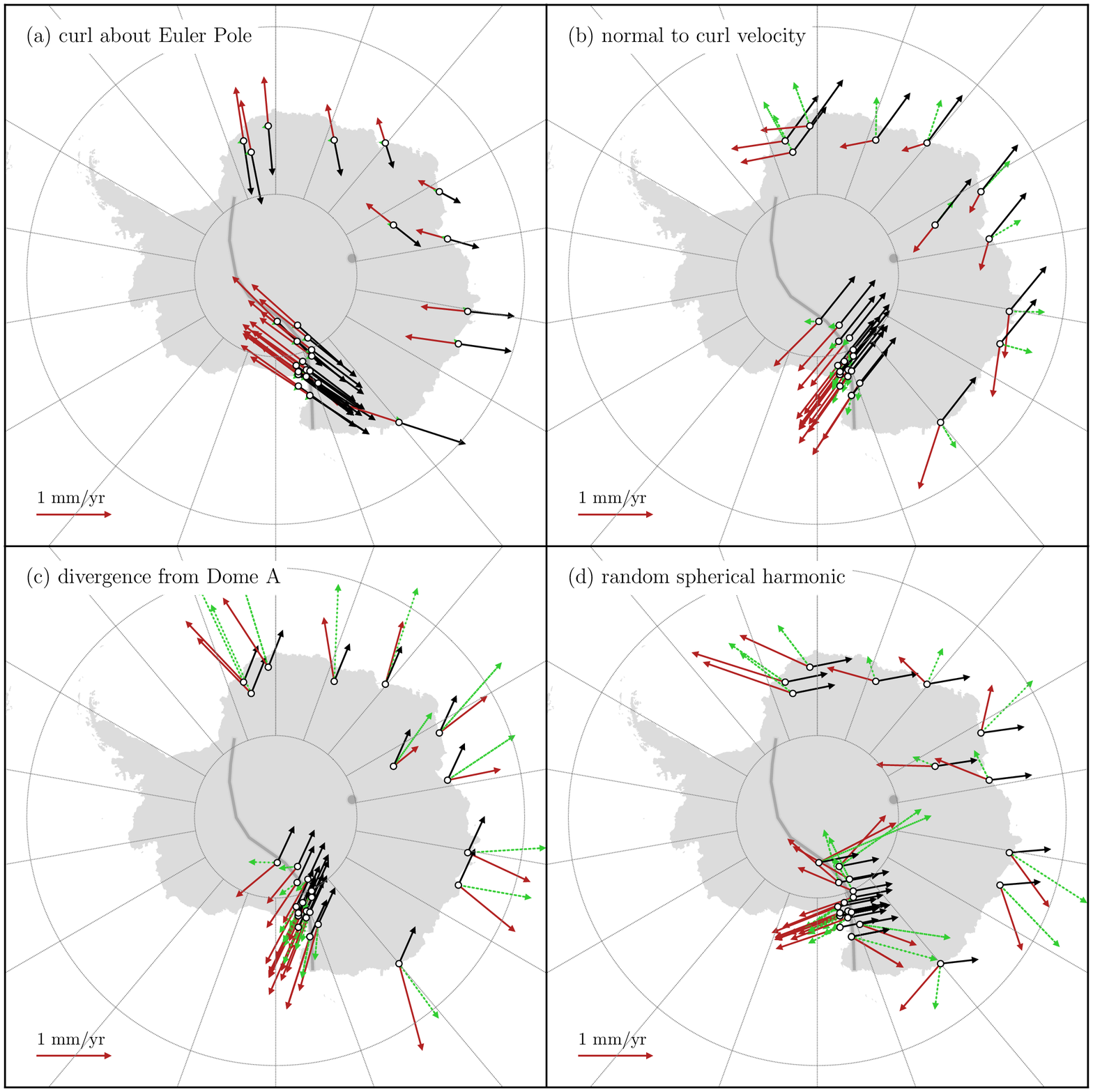}}
\centerline{\includegraphics[width=1.25\columnwidth,trim={105 10 85 10},clip]{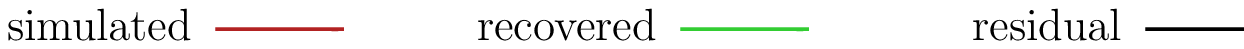}}
\caption{Unweighted regression: a comparison of simulated (solid brown) and recovered (dashed green) velocities for various contrived geometries of a small `GIA' signal in the presence of a larger plate rotation signal at 36 sites in East Antarctica. The difference between the recovered and simulated velocity vectors, i.e. the error in the recovered `GIA' signal, is also shown (black). Velocities are simulated for the `GIA' assuming they are (a) in the direction of curl around the Euler pole; (b) normal to the curl around the Euler pole, (c) in the direction of a divergence field centred on Dome A, (d) randomly generated spherical harmonic functions for the two velocity components. The location of Dome A is shown by a filled grey circle and the approximate location of the Transantarctic Mountains with a curved grey line. {These simulations assume the Euler pole parameters of \citet{Wei-Ping+2009}.}}
\label{fig:Simulated_GIA}
\end{figure*}

When the `GIA' signal is taken from randomly generated spherical harmonic functions for the two velocity components, the `GIA' signal is similarly only half recovered (typical residuals of $\sim$0.46 mm/yr; Table \ref{tab:gia_velocity}) from the simulated vector field, Figure \ref{fig:Simulated_GIA} (d). The recovered `GIA' velocities are again biased by the large number of GPS sites clumped along the Transantarctic Mountains; by construction these have a moderate correlation between their simulated `GIA' velocities, with a sizable component directed approximately parallel to the 100th meridian west. The similar direction of these vectors leads the regression to explain some of the `GIA' field at this location as plate rotation; much smaller residuals are obtained when only including measurements from across the remainder of East Antarctica. By contrast, the typical plate rotation velocities of between 10 and 15 mm/yr have fractional errors of approximately five percent for these last two fields.

\begin{table*}
\caption{Recovered and residual `GIA' velocity vectors (calculated as the difference between the recovered and simulated signal) for the {unweighted} and {areal-}weighted least-squares regression (LSR). The `GIA' velocities are calculated for simulated vectors directed either: (a) following the rotation about the Euler pole (i.e. curl); (b) normal to the rotation about the Euler pole; (c) radially outwards from Dome A on the Antarctic Plateau (i.e. divergence); or (d) from randomly generated spherical harmonic functions for the two velocity components (Figures \ref{fig:Simulated_GIA} and \ref{fig:Weighted_Sim_GIA}). The recovered and residual velocities are also included for two physically-based simulations of the GIA vector field; the tabled values are scaled for a median simulated signal of 1 mm/yr for consistency with our modelling (Figure \ref{fig:Pippa_Sim_GIA}). The median magnitude of the simulated `GIA' vectors in each velocity field is 1 mm/yr. The median and 16th/84th percentiles of the recovered and residual velocity fields are shown. The methods with well recovered signals and minimal residuals are highlighted in bold.}
\small
\centering
\begin{tabular}{l c c c c c c}
\hline
 && \multicolumn{2}{c}{GIA velocity recovered (mm/yr)} && \multicolumn{2}{c}{GIA velocity residuals (mm/yr)} \\
 && {unweighted} LSR & weighted LSR && {unweighted} LSR & weighted LSR \\
\hline
 curl about Euler Pole  && 0 & 0 && $1.00_{\,-0.33}^{\,+0.03}$ & $1.00\pm0.53$ \\
 normal to curl velocity && $0.17_{-0.03}^{+0.35}$ & $0.57\pm0.24$ && $0.86\pm0.01$ & $0.75\pm0.01$ \\
 divergence from Dome A && $0.45_{-0.08}^{+0.83}$ & $\boldsymbol{1.00\pm0.50}$ && $0.57\pm0.01$ & $\boldsymbol{0.35\pm0.01}$ \\
 random spherical harmonics && $0.67_{-0.31}^{+0.47}$ & $\boldsymbol{0.87\pm0.33}$ && $0.46\pm0.21$ & $\boldsymbol{0.31\pm0.19}$ \\
\hline
 4mm grain, wet rheology && $0.26_{-0.09}^{+0.42}$ & $0.42_{-0.32}^{+0.81}$ && $1.16_{-0.22}^{+0.01}$ & $0.83\pm0.43$ \\
 10mm grain, wet rheology && $0.27_{-0.36}^{+0.62}$ & $0.30_{-0.03}^{+1.10}$ && $0.98_{-0.12}^{+0.05}$ & $0.86\pm0.15$ \\
\hline
\end{tabular}
\label{tab:gia_velocity}
\end{table*}

\subsection{{Areal-}weighted regression analysis}
\label{sec:Weighted regression analysis}

The regression analysis undertaken in the previous section for the four simulated `GIA' vector fields is repeated but with each station weighted by the local number density (Figure \ref{fig:Weighted_Sim_GIA}). The `GIA' velocities are poorly recovered for the simulated fields parallel to the rotation about the Euler pole, Figure \ref{fig:Weighted_Sim_GIA} (a). By contrast, the {(weighted) median} residuals for the velocity field normal to the rotation of the Euler pole (b) have reduced in magnitude by 12\% {relative to the unweighted regression} (see Table \ref{tab:gia_velocity}). However, caution should be used in the interpretation of this result as the simulated `GIA' velocities tend to be smaller away from the cluster of sites along the Transantarctic Mountains (i.e. the highest magnitude vectors are down-weighted in the regression, potentially leading to less signal being incorrectly explained as plate rotation).

\begin{figure*}
\centerline{\includegraphics[width=1.25\columnwidth,trim={105 105 85 100},clip]{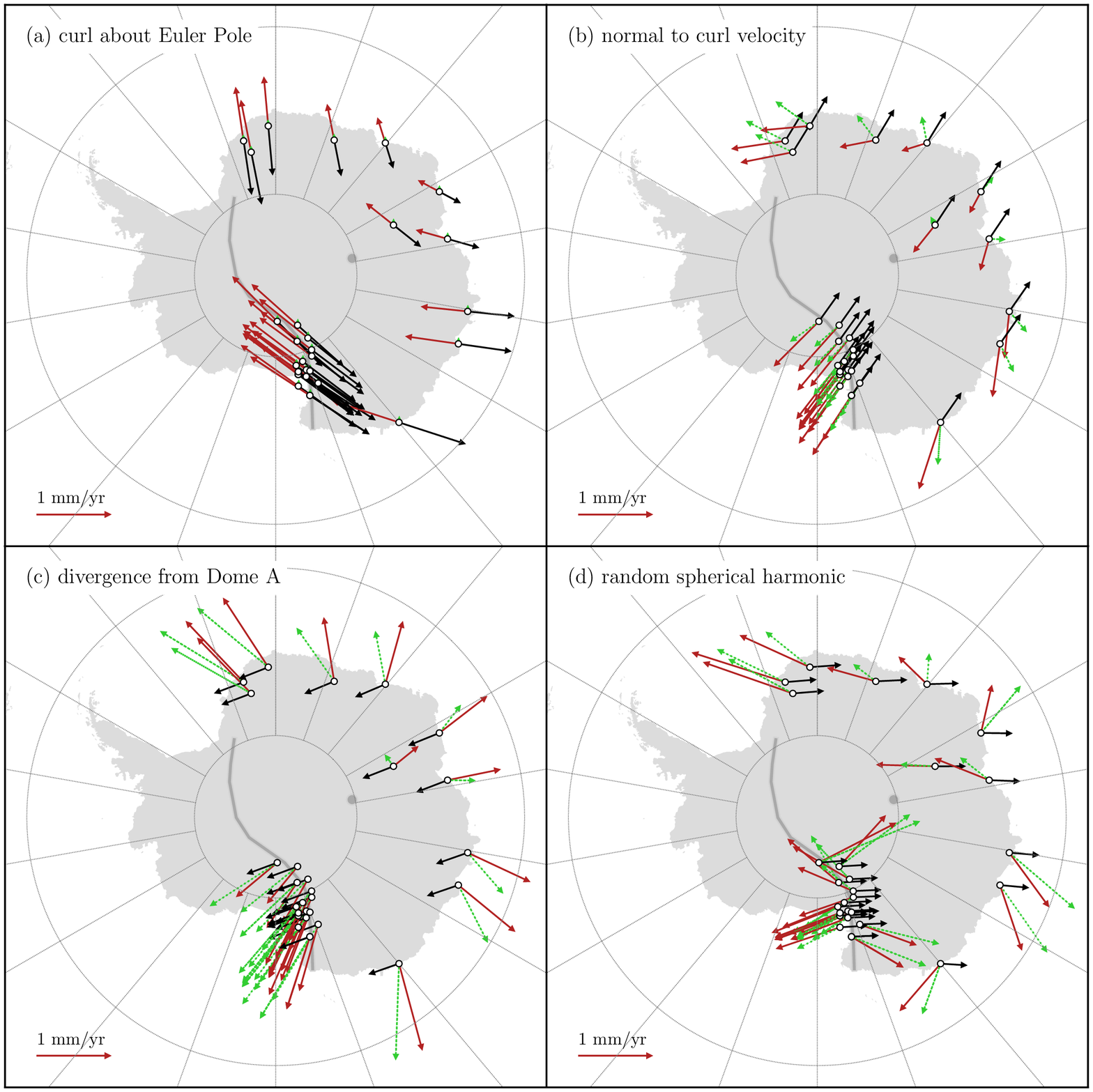}}
\centerline{\includegraphics[width=1.25\columnwidth,trim={105 10 85 10},clip]{Figures/legend.eps}}
\caption{{Areal-}weighted regression for the simulated `GIA' signals investigated in Figure \ref{fig:Simulated_GIA}; see caption to that figure for further description. The East Antarctic stations are weighted by the inverse of the local number density, as shown in Figure \ref{fig:Number_densities}.}
\label{fig:Weighted_Sim_GIA}
\end{figure*}

The `GIA' signal simulated for velocities aligned along the divergence field centred on Dome A (c) are now largely recovered (magnitudes almost completely recovered), with their residuals reduced by 38\%. These vectors are now directed approximately parallel to the 110th meridian west (previously parallel to the 20th meridian east) since the distribution of stations only form a partial circle about Dome A; i.e. none of the `GIA' vectors point towards the western hemisphere so the regression fits part of the vector field (towards the 70th meridian east) as plate rotation.
The `GIA' signal is also mostly recovered when the simulated `GIA' vector field at each station is taken from randomly generated spherical harmonics for the two velocity components (d). Specifically, the weighted median recovered `GIA' velocity vector is within 13\% of the simulated magnitude, whilst the median residual (including errors in direction) is 0.31 mm/yr, or approximately 31\%. These residuals lead to errors in the plate rotation velocities (typically 10-15 mm/yr) of two to three percent.

The {areal-}weighted regression more accurately recovers the simulated `GIA' velocities, i.e. separates the plate rotation and `GIA' signals, compared to the {unweighted} regression analysis by avoiding the introduction of bias, notably from the 25 correlated GPS stations along the Transantarctic Mountains. Neither technique can separate the `GIA' and plate rotation signals when the GIA vectors are highly correlated across the continent. However, if the `GIA' velocities at each station are sufficiently uncorrelated the {unweighted} regression recovers velocities with residuals greater than 45\% of the simulated magnitude whilst the weighted regression has residuals less than 35\% of the magnitude of the original vector field. The recovered `GIA' velocities, and their residuals, obtained for each simulated vector field and using both regressions are shown in Table \ref{tab:gia_velocity}.

\begin{figure*}
\centerline{\includegraphics[width=1.25\columnwidth,trim={105 50 85 50},clip]{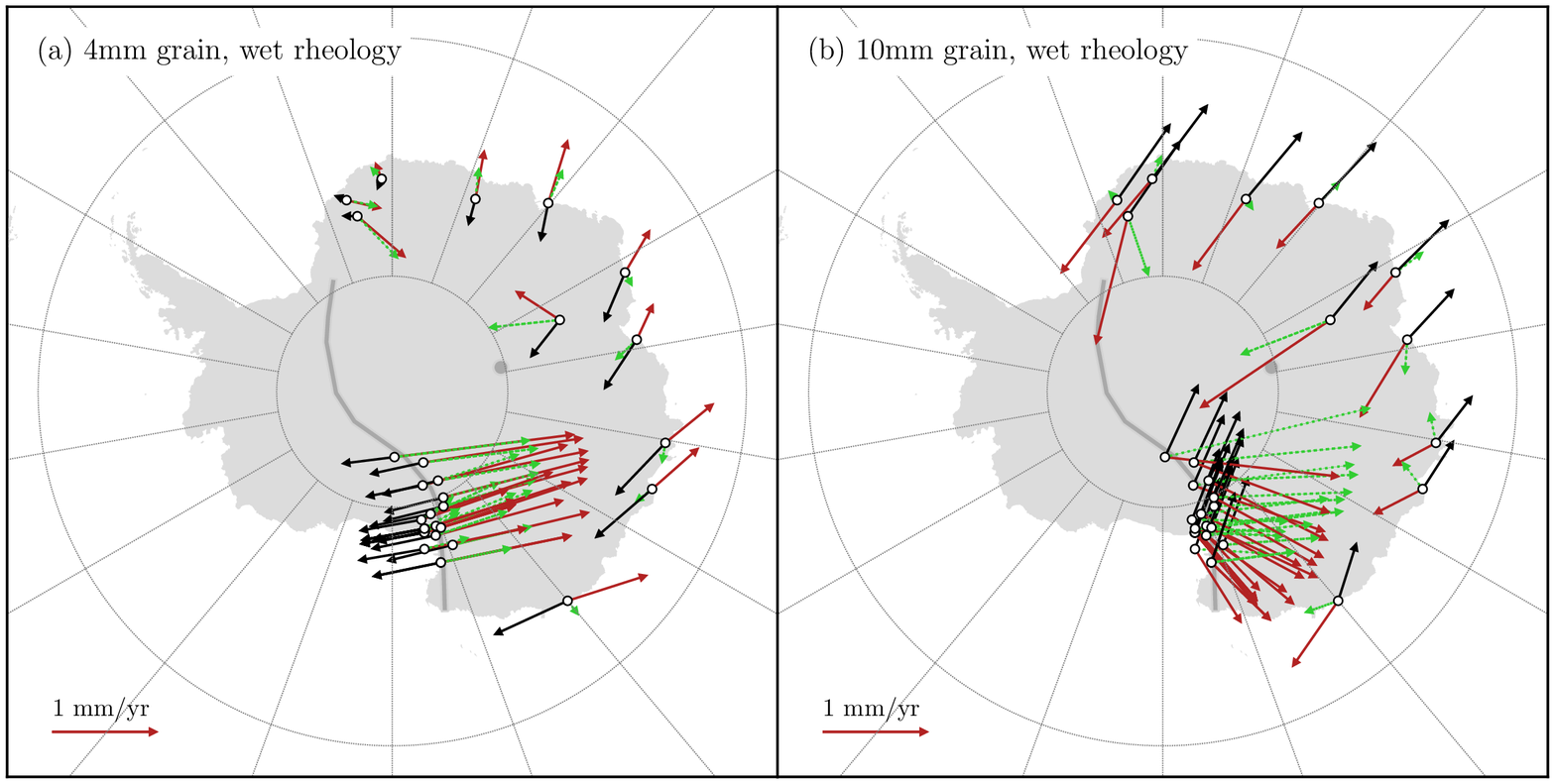}}
\centerline{\includegraphics[width=1.25\columnwidth,trim={105 10 85 10},clip]{Figures/legend.eps}}
\caption{Effectiveness of the {areal-}weighted regression on two physically-based simulations of the GIA signal in East Antarctica. GIA horizontal velocities are simulated using the W12 ice model based on the S40RTS seismic model for a wet rheology assuming: (a) 4mm grain size, and (b) 10mm grain size. {These simulations assume the Euler pole parameters of \citet{Wei-Ping+2009}.} See caption to Figure \ref{fig:Simulated_GIA} for further description.}
\label{fig:Pippa_Sim_GIA}
\end{figure*}

The ability of the weighted regression to recover physically-based GIA simulations is also tested (Table \ref{tab:gia_velocity} and Figure \ref{fig:Pippa_Sim_GIA}). {Two of these physically-based GIA models (discussed in Section \ref{sec:Areal weighting}) are shown in Figure \ref{fig:Pippa_Sim_GIA}; these both assume the S40RTS seismic model of \citet{Ritsema+2011}, a wet rheology, and either 4 or 10 mm grain sizes. Importantly, the direction of the GIA vectors at individual stations varies by as much as 180 degrees between these two models so they cannot be used to validate the output of our regressions on actual data on a station-by-station basis.}
The residual velocity vectors for both models are approximately 80\% of the simulated magnitude, whilst the recovered signal is between 30 to 42\%; i.e. these simulated signals would be poorly recovered and not detectable above the systematic uncertainties. This primarily results from the large asymmetry in GIA magnitudes between the Ross Sea stations and those in coastal East Antarctica. However, these physically-based models all predict a median GIA horizontal velocity in East Antarctica of between 0.6 and 1.15 mm/yr (typically 0.9 mm/yr) with no signals exceeding 2 mm/yr. Measurements of the actual GIA velocity field (using our regression analysis) can rule out these asymmetric GIA fields if the observed field strength exceeds that which can theoretically be recovered.

\section{Separation of velocity components from observed data}
\label{sec:antarctic}

The Monte Carlo regression analysis technique described in Section \ref{sec:Regression analysis} is applied to the measured horizontal velocities from 36 GPS station sites across East Antarctica. This application to \emph{actual} data aims to inform the separation of signals relating to the plate rotation and local components, attributed to GIA, and we provide an appraisal of the uncertainty in the resulting velocities.

The best fitting values for the rotation rate of the Antarctic Plate and the location of the Euler pole are listed in Table \ref{tab:plate_rotation} for both the {unweighted} regression and {areal-}weighted regression (weighting each station by the local number density of GPS stations; see Section \ref{sec:Weighted regression analysis}). {The corresponding variance-covariance matrices are represented graphically in Figure \ref{fig:covariances}.} The angular velocity, and the latitude and longitude of the Euler pole, fitted using these two techniques are inconsistent with each other at the $2\sigma$ level differing by 2.5 degrees in latitude, one degree in longitude, and approximately 10\% in rotation rate. Our plate motion fitted using the {unweighted} LSR is also broadly in agreement with the findings of other authors in the literature \citep[e.g.][these are also included in Table \ref{tab:plate_rotation} {and shown in Figure \ref{fig:euler_poles}}]{Larson+1997, Sella+2002, Altamimi+2002, Wei-Ping+2009, Argus+2014, Ghavri+2015}, though all parameters unsurprisingly differ more significantly when fitted using the weighted regression. {Specifically, the weighted Euler pole lies 380-750 km further off the East Antarctic coast whilst its rotation rate is 13-17\% slower. {The motion of the Antarctic Plate is also parameterised in Table \ref{tab:plate_rotation} in terms of rotation rates about the three cartesian coordinate axes; the variance-covariance matrices are again shown in Figure \ref{fig:covariances} and tabulated in Supplement C. The $z$-component of the rotation rate in our areal-weighted LSR is again 13-15\% slower than for both the unweighted LSR and the fits of other authors \citep[e.g.][]{Altamimi+2012}.} These results suggest that the Euler pole parameters derived by these other authors may be somewhat biased by an asymmetric geometric distribution of stations (see Section \ref{sec:Station configuration}).} 

\begin{table*}
\caption{Rotation rates and Euler poles for the Antarctic Plate based on {{observational} data} using the {areal-}weighted and {unweighted} LSR, and a comparison to results of other studies. {The Euler pole parameters are further calculated using the {areal-}weighted LSR but assuming different correlation length-scales.} The columns are: Euler pole latitude, $\theta_\epsilon$; and longitude, $\phi_\epsilon$; angular velocity {cartesian components}, $\omega_{\rm x}$, $\omega_{\rm y}$ and $\omega_{\rm z}$ {(anticlockwise about x, y and z axes is positive)}; {and Euler pole angular velocity, $\omega$ (anticlockwise about pole is positive)}. Statistical uncertainties are included at the 1$\sigma$ confidence level.} 
\small
\centering
\begin{tabular}{l c c c c c c}
\hline
 & Latitude & Longitude & \multicolumn{4}{c}{Angular velocity}  \\
 & $\theta_\epsilon$ ($^\circ$N) & $\phi_\epsilon$ ($^\circ$E) & $\omega_{\rm x}$ (mas/a) & $\omega_{\rm y}$ (mas/a) & $\omega_{\rm z}$ (mas/a) & $\omega$ ($^\circ/\rm Ma$) \\
\hline
  weighted LSR (5$^\circ$ length-scale)  & $55.04_{-0.72}^{+0.65}$ & $-126.17_{-0.91}^{+0.93}$ & \!\!$-0.232\pm0.007$\! & \!\!$-0.318\pm0.008$\! & $0.563\pm0.022$ & $0.191\pm0.006$ \\
  {unweighted} LSR & $57.59_{-0.37}^{+0.38}$ & $-127.15_{-0.34}^{+0.35}$ & \!\!$-0.253\pm0.002$\! & \!\!$-0.334\pm0.002$\! & $0.661\pm0.009$ & $0.217\pm0.002$ \\
\hline
%\citet{Ghavri+2015}  & $0.1814\pm0.0051$ & $57.74\pm0.41$ & $-126.90\pm0.40$ \\
  \citet{Argus+2014} & $59.876$ & $-127.277$ & -- & -- & -- & $0.2178$ \\
  \citet{Altamimi+2012} & -- & -- & \!\!$-0.252\pm0.008$\! & \!\!$-0.302\pm0.006$\! & $0.643\pm0.009$ & $0.209\pm0.003$ \\
  \citet{Wei-Ping+2009} & $58.69\pm0.3$ & $-128.29\pm0.4$ & -- & -- & -- & $0.224\pm0.01$ \\
  \citet{Altamimi+2002} & $61.8$ & $-125.6$ & -- & -- & -- & $0.23\pm0.02$ \\
  \citet{Sella+2002} & $58.5\pm1.6$ & $-134\pm1$ & \!\!$-0.295\pm0.001$\! & \!\!$-0.306\pm0.001$\! & $0.693\pm0.006$ & $0.23\pm0.01$ \\
%  \citet{Larson+1997}  & $0.24\pm0.03$ & $60.5\pm6.6$ & $-125.7\pm3.6$ \\
\hline
  weighted LSR (2.5$^\circ$ length-scale) & $55.95_{-0.53}^{+0.51}$ & $-125.55_{-0.68}^{+0.68}$ & \!\!$-0.234\pm0.005$\! & \!\!$-0.327\pm0.005$\! & $0.596\pm0.015$ & $0.199\pm0.004$ \\
  weighted LSR (7.5$^\circ$ length-scale) & $54.90_{-0.90}^{+0.82}$ & $-126.14_{-0.98}^{+1.05}$ & \!\!$-0.232\pm0.008$\! & \!\!$-0.319\pm0.009$\! & $0.561\pm0.028$ & $0.191\pm0.007$ \\
  weighted LSR (10$^\circ$ length-scale) & $55.24_{-0.86}^{+0.82}$ & $-125.72_{-0.95}^{+0.98}$ & \!\!$-0.230\pm0.007$\! & \!\!$-0.320\pm0.009$\! & $0.567\pm0.026$ & $0.192\pm0.007$ \\
\hline
\end{tabular}
\label{tab:plate_rotation}
\end{table*}

\begin{figure*}
\centerline{\includegraphics[width=1.6\columnwidth,trim={0 15 35 30},clip]{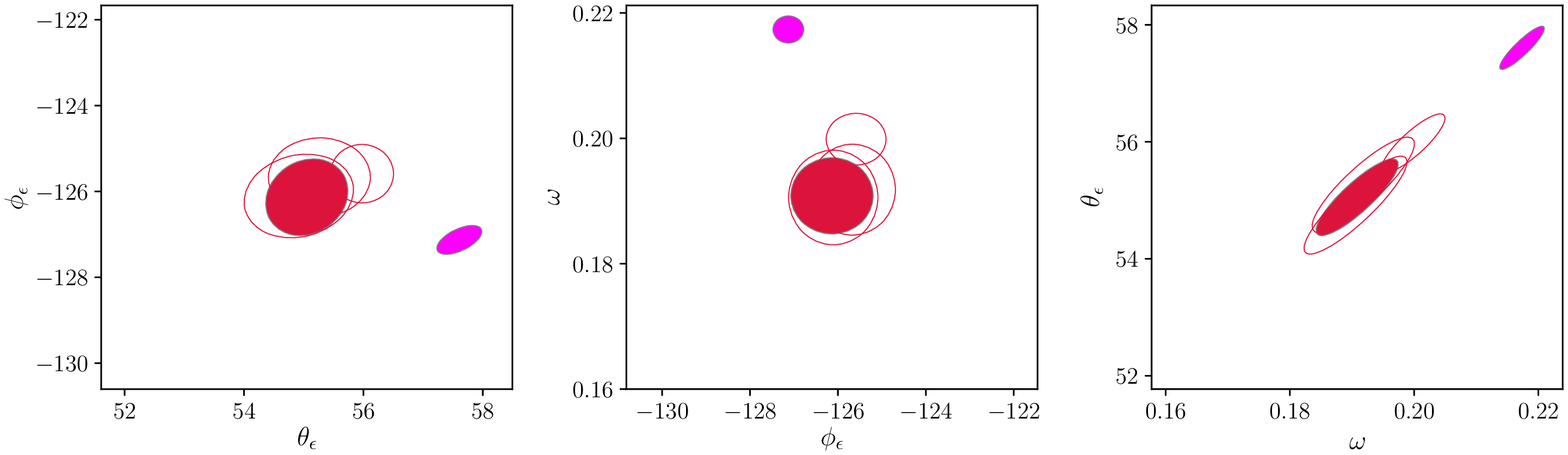}}\vspace{3pt}
\centerline{\includegraphics[width=1.6\columnwidth,trim={0 15 35 30},clip]{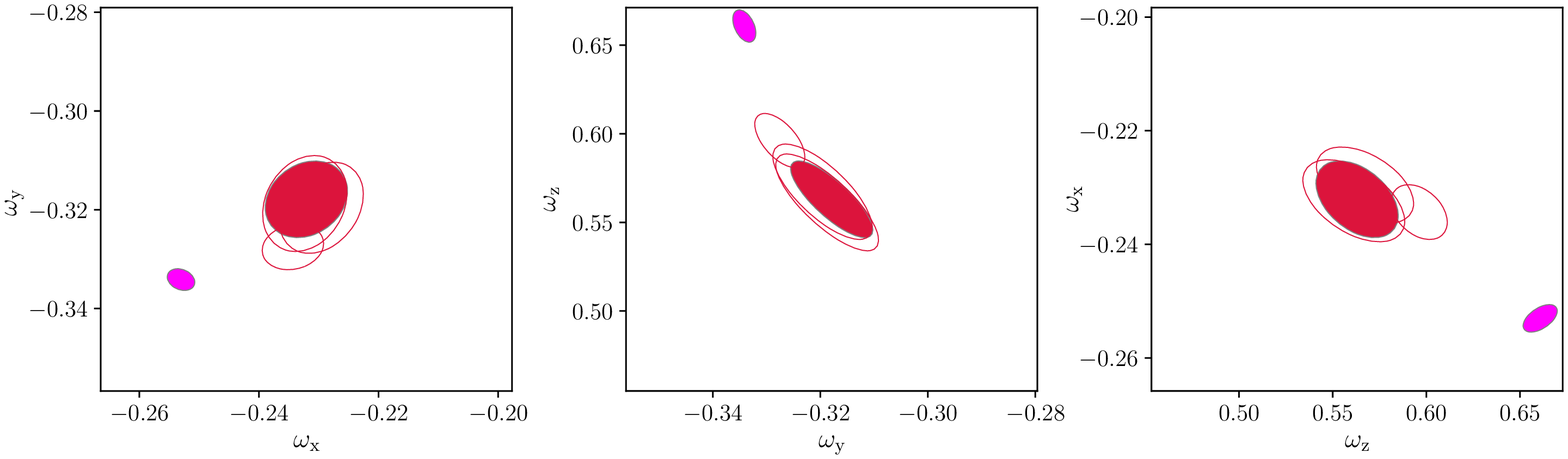}}
\caption{{Variance-covariance matrices for the fitted motion of the Antarctic Plate expressed as confidence ellipses at the 1$\sigma$ level. The variance-covariance ellipses for motion parameterised in terms of rotation about an Euler pole (i.e. $\theta_\epsilon$, $\phi_\epsilon$, $\omega$) are shown in the top three sub-figures; the variance-covariance ellipses for motion parameterised in terms of rotation about the three cartesian axes (i.e. $\omega_{\rm x}$, $\omega_{\rm y}$, $\omega_{\rm z}$) are in the bottom three sub-figures. Each sub-figure includes fits for the {areal-}weighted (red) and unweighted (pink) regressions; see caption to Figure \ref{fig:euler_poles}.}}
\label{fig:covariances}
\end{figure*}

{The stability of the Euler pole parameters to moderate changes in the correlation length-scale from our assumed value of 5 degrees is also explored (Figures \ref{fig:covariances} and \ref{fig:euler_poles}, and Table \ref{tab:plate_rotation}); we find both the location and rotation rate are consistent at the $1\sigma$ level for length-scales between at least 2.5 and 10 degrees. The behaviour of the areal-weighted regression converges to that of the unweighted regression as the length-scale shrinks to comprise only a single station, or enlarges to contain all stations.}

\begin{figure*}
\centerline{\includegraphics[width=1.25\columnwidth,trim={105 50 105 50},clip]{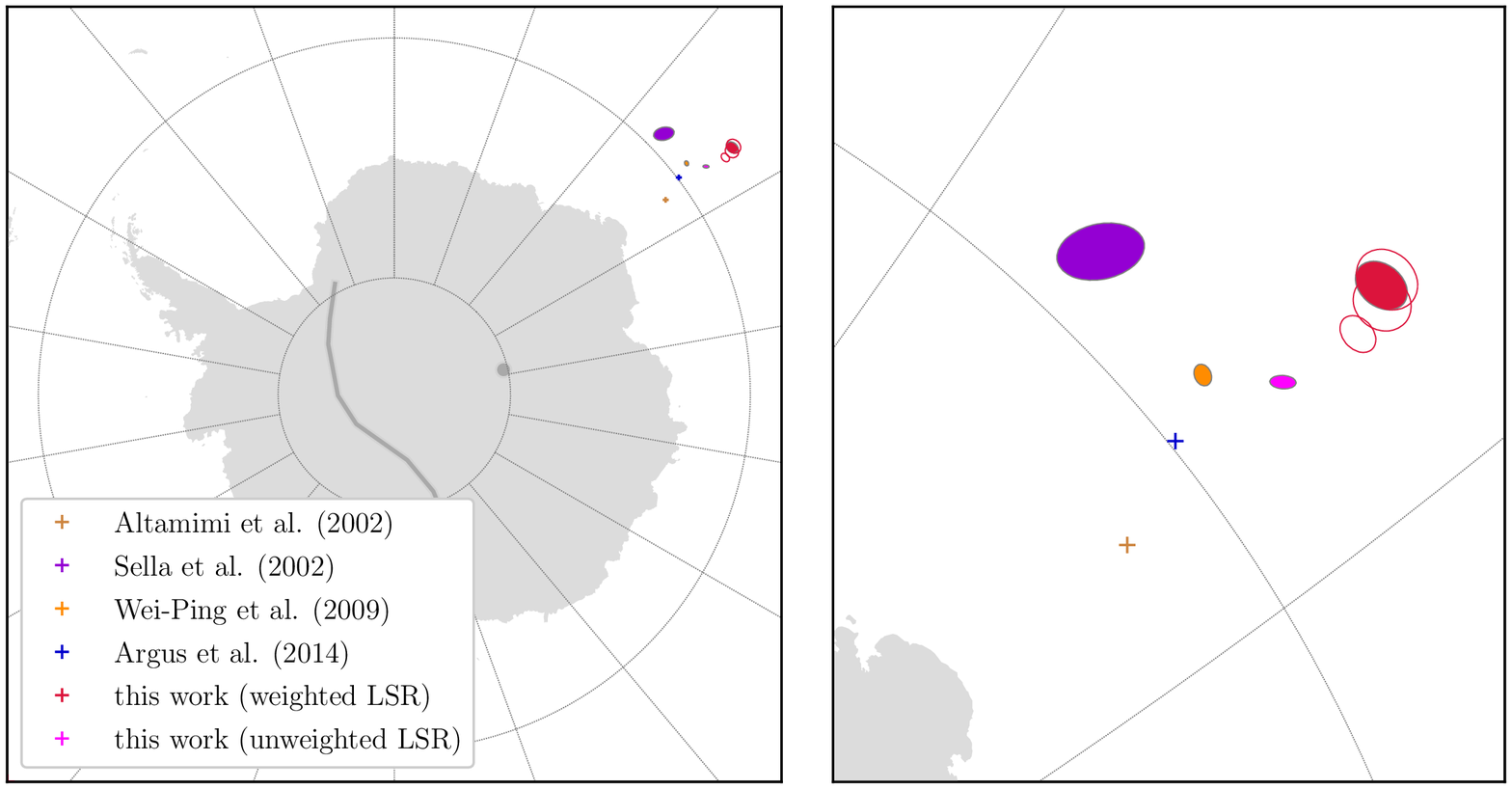}}
\caption{{Location of the (antipodal) Euler pole of the Antarctic Plate calculated from the {observational} data used in this work for both the {areal-}weighted (red) and unweighted (pink) regressions. Confidence ellipses are shown at the $1\sigma$ level including the covariance between the latitude and longitude. The unfilled red ellipses show the location of the Euler pole for the {areal-}weighted LSR assuming different correlation length-scales (see Table \ref{tab:plate_rotation}). The location of the Euler pole derived by other authors in the literature is shown for comparison (see figure caption); these measurements do not report covariances whilst those with no quoted uncertainties are shown as crosses.}}
\label{fig:euler_poles}
\end{figure*}

The local velocity field estimated from the measured horizontal velocities upon subtracting the best fitting plate rotation, based on both {unweighted} and {areal-}weighted regression, is shown in Figure \ref{fig:Antarctic_GIA}. The {local} velocities for individual stations using the {areal-}weighted regression are included in Table \ref{tab:GIA_velocities}. The velocities along the Transantarctic Mountains are (incorrectly) biased in the {unweighted} fit to have approximately random directions (since their correlated {local} velocities are fitted as plate rotation), whereas the weighted regression consistently fits velocities pointing towards the Ross Sea. The velocity field across the rest of East Antarctica also shows a slight preference to be directed outwards towards the coast for both fitting techniques. The horizontal velocities due to the rotation of the Antarctic Plate are also shown in Figure \ref{fig:Antarctic_GIA}, however these vector fields are largely indistinguishable from each other as expected \citep[e.g. relative magnitudes of the fields in][]{Peltier+2015}.

{The (weighted) median magnitude of the local velocity vectors found using the areal-weighted regression is $0.80_{-0.06}^{+0.09}$ mm/yr, whilst the maximum velocity is $3.14\pm0.87$ mm/yr. These are comparable to or higher than expected by physically-based simulations of the GIA signal in the range 0.6-1.15 mm/yr (Section \ref{sec:Areal weighting}). The majority of the GIA signal measured by the {actual} GPS stations must therefore be recovered by the signal separation technique; i.e. if recovered local velocities only capture a small fraction of the actual signal we would expect GIA velocities much greater than 0.8 mm/yr. These high levels of signal recovery are only possible if the local velocities are uncorrelated on large length-scales (e.g. following a divergent pattern or random spherical harmonics, see Section \ref{sec:Weighted regression analysis}).
The systematic error in fitting the local component of the signal using our regression analysis is therefore at most 0.28 mm/yr {(i.e. approximately 35\% for divergence from Dome A, as discussed in Section \ref{sec:Weighted regression analysis})}, though is likely less if the signal is less correlated than this worst case scenario. We assume the systematic uncertainty in our method is that simulated for the `more general' random spherical harmonics of 0.25 mm/yr (i.e. approximately 31\%).}

\begin{figure*}
\centerline{\includegraphics[width=1.25\columnwidth,trim={105 50 85 50},clip]{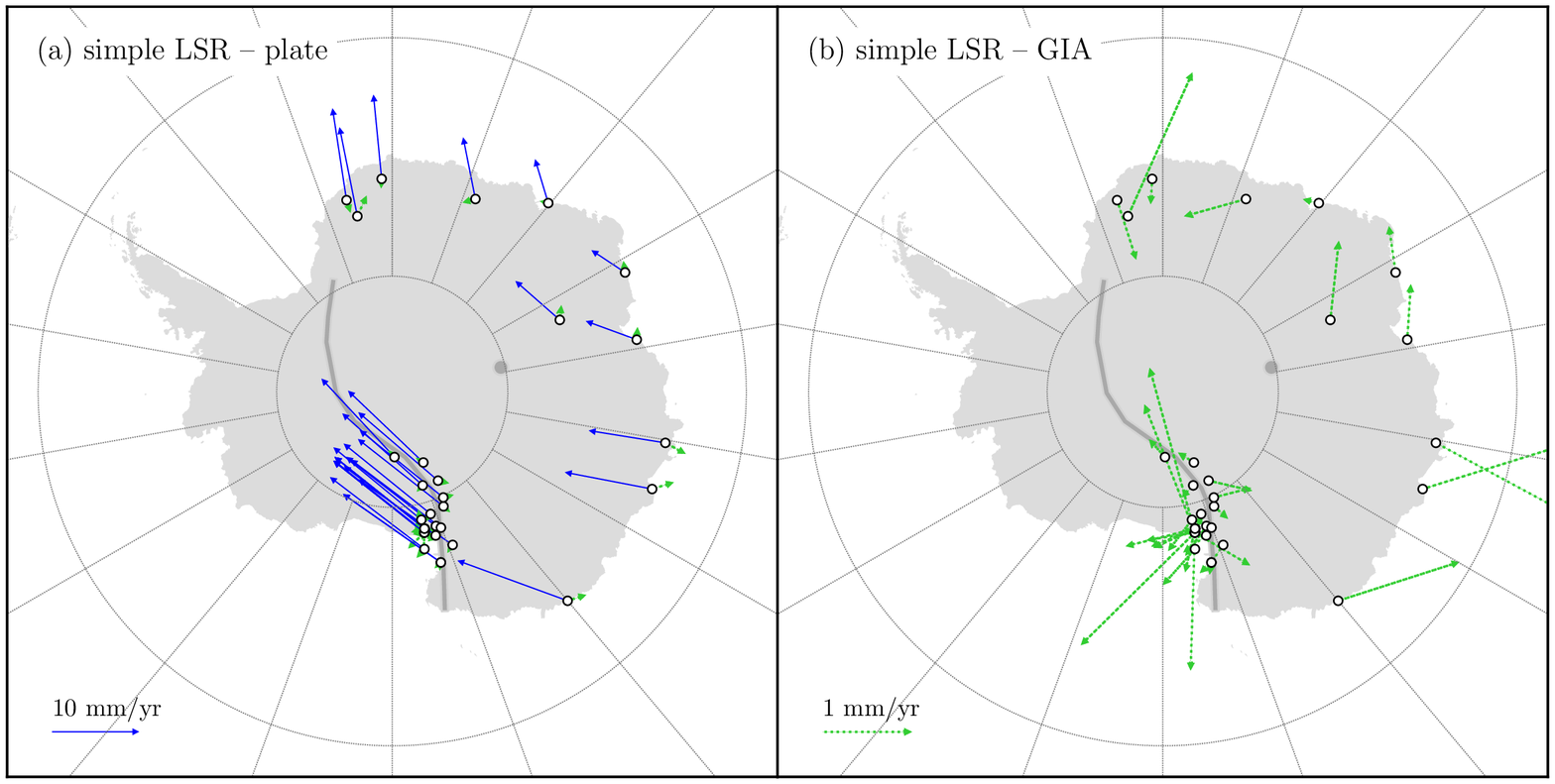}}\vspace{0.15cm}
\centerline{\includegraphics[width=1.25\columnwidth,trim={105 50 85 50},clip]{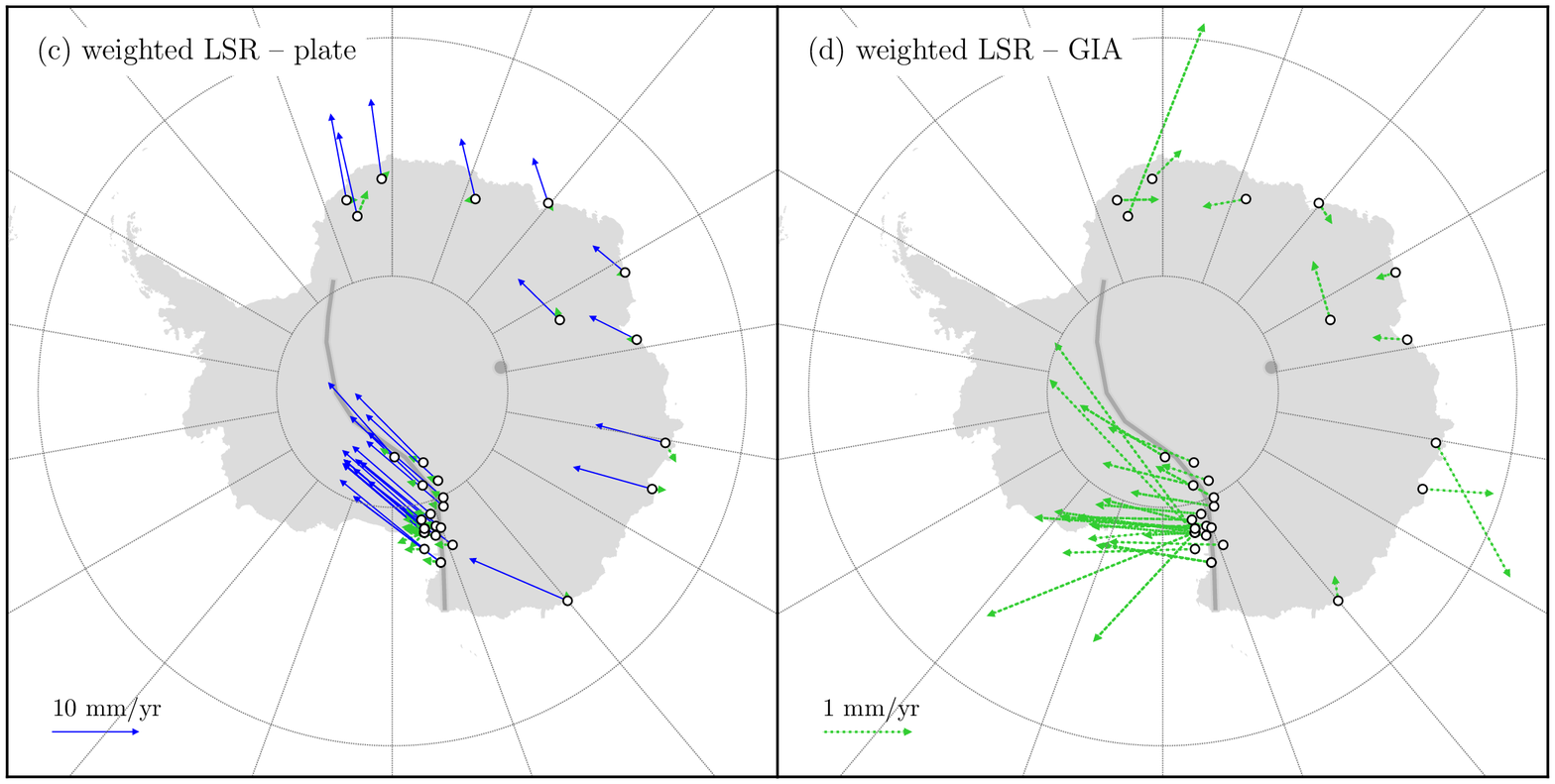}}
\caption{Best fitting horizontal velocities due to the rotation of the Antarctic Plate (solid blue) and GIA (dashed green) at sites around the Antarctic continent. The fits are made using only the measurements from East Antarctic stations (black unfilled circles). Fits using the {unweighted} regression are shown in (a) and (b), whilst (c) and (d) are for the {areal-}weighted regression.}
\label{fig:Antarctic_GIA}
\end{figure*}

The uncertainties on the {local} velocities provided in Table \ref{tab:GIA_velocities} include both measurement uncertainties and, as above, the systematic errors expected from the {signal separation} analysis. The measurement uncertainties in the {local} velocities, calculated using the Monte Carlo simulation for the weighted regression, are also shown in Figure \ref{fig:Antarctic_GIA_errors}. The weighted regression produces an approximately uniform distribution of measurements uncertainties in the GIA horizontal velocities. Non-zero velocities (assuming both measurement and sysmtematic uncertainties) are fitted for 72\% of the GPS stations at the $1\sigma$ level and 58\% and for the higher $2\sigma$ confidence level. A higher fraction of stations have significant east-west components of their {local} velocities: 81\% at the 1$\sigma$ level and 69\% for 2$\sigma$ confidence. The GPS stations with 2$\sigma$ GIA velocity detections are mostly located along the Transantarctic Mountains, though the Svea Research Station is the notable exception. A368 and BHIL have a single velocity component significant at the 2$\sigma$ level.

\begin{table*}
\caption{{Local} velocity vectors at each of the 36 GPS sties in East Antarctica. The station code is listed in the first column, the latitude and longitude in the second and third column, the north, east and radial components of the vectors in the fourth through sixth columns, and the direction of the best fit vector is included in the final column. Vector component velocities shown in bold are significant at the 2$\sigma$ confidence level.}
\small
\centering
\renewcommand{\arraystretch}{0.8}
\begin{tabular}{l c c c c c c}
\hline
Station &	Latitude ($^\circ$N) &	Longitude ($^\circ$E) &	\multicolumn{3}{c}{GIA velocity components (mm/yr)} &	Direction	\\
 & & & north &	east &	magnitude \\
\hline
A368 & -74.2906 & 66.7922 & $0.08\pm0.82$ & $\boldsymbol{-0.73\pm0.32}$ & $0.74\pm0.89$ & 277$^\circ$	\\
ABOA & -73.0438 & -13.4072 & $-0.10\pm0.43$ & $0.46\pm0.42$ & $0.47\pm0.64$ & 102$^\circ$	\\
BHIL & -66.2510 & 100.5990 & $1.26\pm0.81$ & $\boldsymbol{1.49\pm0.60}$ & $1.95\pm1.06$ & 51$^\circ$	\\
BRIP & -75.7957 & 158.4691 & $-0.57\pm0.39$ & $\boldsymbol{1.34\pm0.43}$ & $\boldsymbol{1.46\pm0.61}$ & 113$^\circ$	\\
BURI & -79.1475 & 155.8941 & $-0.57\pm0.36$ & $\boldsymbol{0.88\pm0.39}$ & $1.05\pm0.56$ & 123$^\circ$	\\
CAS1 & -66.2834 & 110.5197 & $0.81\pm0.42$ & $-0.24\pm0.29$ & $0.85\pm0.54$ & 344$^\circ$	\\
CONG & -77.5346 & 167.0855 & $0.46\pm0.48$ & $\boldsymbol{2.93\pm0.57}$ & $\boldsymbol{2.96\pm0.79}$ & 82$^\circ$	\\
COTE & -77.8059 & 161.9979 & $-0.45\pm0.37$ & $\boldsymbol{1.40\pm0.42}$ & $\boldsymbol{1.47\pm0.59}$ & 108$^\circ$	\\
DAV1 & -68.5773 & 77.9726 & $-0.35\pm0.43$ & $-0.10\pm0.26$ & $0.36\pm0.53$ & 196$^\circ$	\\
DEVI & -81.4767 & 161.9771 & $-0.62\pm0.38$ & $\boldsymbol{1.01\pm0.40}$ & $\boldsymbol{1.19\pm0.58}$ & 122$^\circ$	\\
DUM1 & -66.6651 & 140.0019 & $-0.20\pm0.44$ & $-0.12\pm0.45$ & $0.24\pm0.66$ & 212$^\circ$	\\
E1G2 & -77.5304 & 167.1396 & $\boldsymbol{1.11\pm0.52}$ & $\boldsymbol{1.60\pm0.50}$ & $\boldsymbol{1.94\pm0.77}$ & 56$^\circ$	\\
FIE0 & -76.1446 & 168.4235 & $-0.30\pm0.50$ & $\boldsymbol{1.69\pm0.54}$ & $\boldsymbol{1.72\pm0.78}$ & 100$^\circ$	\\
FLM5 & -77.5327 & 160.2714 & $\boldsymbol{-0.81\pm0.38}$ & $\boldsymbol{1.19\pm0.40}$ & $\boldsymbol{1.43\pm0.59}$ & 124$^\circ$	\\
FTP4 & -78.9277 & 162.5647 & $-0.52\pm0.36$ & $\boldsymbol{1.26\pm0.41}$ & $\boldsymbol{1.36\pm0.57}$ & 112$^\circ$	\\
HOOZ & -77.5316 & 166.9326 & $-0.67\pm0.44$ & $\boldsymbol{1.73\pm0.52}$ & $\boldsymbol{1.85\pm0.72}$ & 111$^\circ$	\\
IGGY & -83.3072 & 156.2498 & $\boldsymbol{-0.85\pm0.40}$ & $\boldsymbol{0.82\pm0.38}$ & $\boldsymbol{1.18\pm0.59}$ & 136$^\circ$	\\
LEHG & -77.5106 & 167.1411 & $\boldsymbol{-2.87\pm0.68}$ & $\boldsymbol{1.26\pm0.48}$ & $\boldsymbol{3.14\pm0.87}$ & 156$^\circ$	\\
LWN0 & -81.3459 & 152.7316 & $-0.41\pm0.43$ & $0.40\pm0.38$ & $0.58\pm0.60$ & 136$^\circ$	\\
LXAA & -71.9466 & 23.3462 & $-0.28\pm0.42$ & $-0.40\pm0.46$ & $0.49\pm0.67$ & 235$^\circ$	\\
MACG & -77.5326 & 167.2466 & $-0.10\pm0.37$ & $0.59\pm0.45$ & $0.60\pm0.62$ & 99$^\circ$	\\
MAW1 & -67.6048 & 62.8706 & $-0.17\pm0.39$ & $-0.03\pm0.23$ & $0.17\pm0.48$ & 190$^\circ$	\\
MCM4 & -77.8384 & 166.6693 & $-0.64\pm0.44$ & $\boldsymbol{2.02\pm0.49}$ & $\boldsymbol{2.11\pm0.70}$ & 108$^\circ$	\\
MCMC & -77.8384 & 166.6693 & $-0.56\pm0.34$ & $\boldsymbol{1.67\pm0.41}$ & $\boldsymbol{1.76\pm0.55}$ & 109$^\circ$	\\
MCMD & -77.8384 & 166.6693 & $-0.55\pm0.35$ & $\boldsymbol{1.65\pm0.43}$ & $\boldsymbol{1.74\pm0.58}$ & 108$^\circ$	\\
MIN0 & -78.6503 & 167.1637 & $-0.36\pm0.42$ & $\boldsymbol{1.42\pm0.53}$ & $\boldsymbol{1.46\pm0.72}$ & 104$^\circ$	\\
NAUS & -77.5219 & 167.1472 & $\boldsymbol{-2.41\pm0.57}$ & $\boldsymbol{1.43\pm0.45}$ & $\boldsymbol{2.80\pm0.77}$ & 149$^\circ$	\\
RAMG & -84.3384 & 178.0471 & $-0.69\pm0.36$ & $\boldsymbol{1.06\pm0.38}$ & $\boldsymbol{1.27\pm0.55}$ & 123$^\circ$	\\
ROB4 & -77.0345 & 163.1901 & $-0.57\pm0.35$ & $\boldsymbol{1.39\pm0.42}$ & $\boldsymbol{1.51\pm0.58}$ & 112$^\circ$	\\
SCTB & -77.8490 & 166.7579 & $-0.18\pm0.42$ & $\boldsymbol{1.36\pm0.52}$ & $1.37\pm0.71$ & 98$^\circ$	\\
SVEA & -74.5760 & -11.2252 & $\boldsymbol{2.28\pm0.60}$ & $\boldsymbol{1.47\pm0.41}$ & $\boldsymbol{2.71\pm0.77}$ & 34$^\circ$	\\
SYOG & -69.0070 & 39.5837 & $-0.07\pm0.30$ & $0.21\pm0.38$ & $0.22\pm0.50$ & 108$^\circ$	\\
TNB1 & -74.6988 & 164.1029 & $-0.61\pm0.42$ & $\boldsymbol{1.36\pm0.46}$ & $\boldsymbol{1.49\pm0.66}$ & 114$^\circ$	\\
TNB2 & -74.6990 & 164.1028 & $-0.60\pm0.41$ & $\boldsymbol{1.28\pm0.49}$ & $\boldsymbol{1.42\pm0.68}$ & 115$^\circ$	\\
VESL & -71.6738 & -2.8418 & $0.31\pm0.43$ & $0.33\pm0.38$ & $0.45\pm0.60$ & 47$^\circ$	\\
WHN0 & -79.8457 & 154.2202 & $-0.65\pm0.44$ & $0.48\pm0.39$ & $0.80\pm0.62$ & 143$^\circ$	\\
\hline
\end{tabular}
\label{tab:GIA_velocities}
\end{table*}

\begin{figure*}
\centerline{\includegraphics[width=1.25\columnwidth,trim={105 50 105 50},clip]{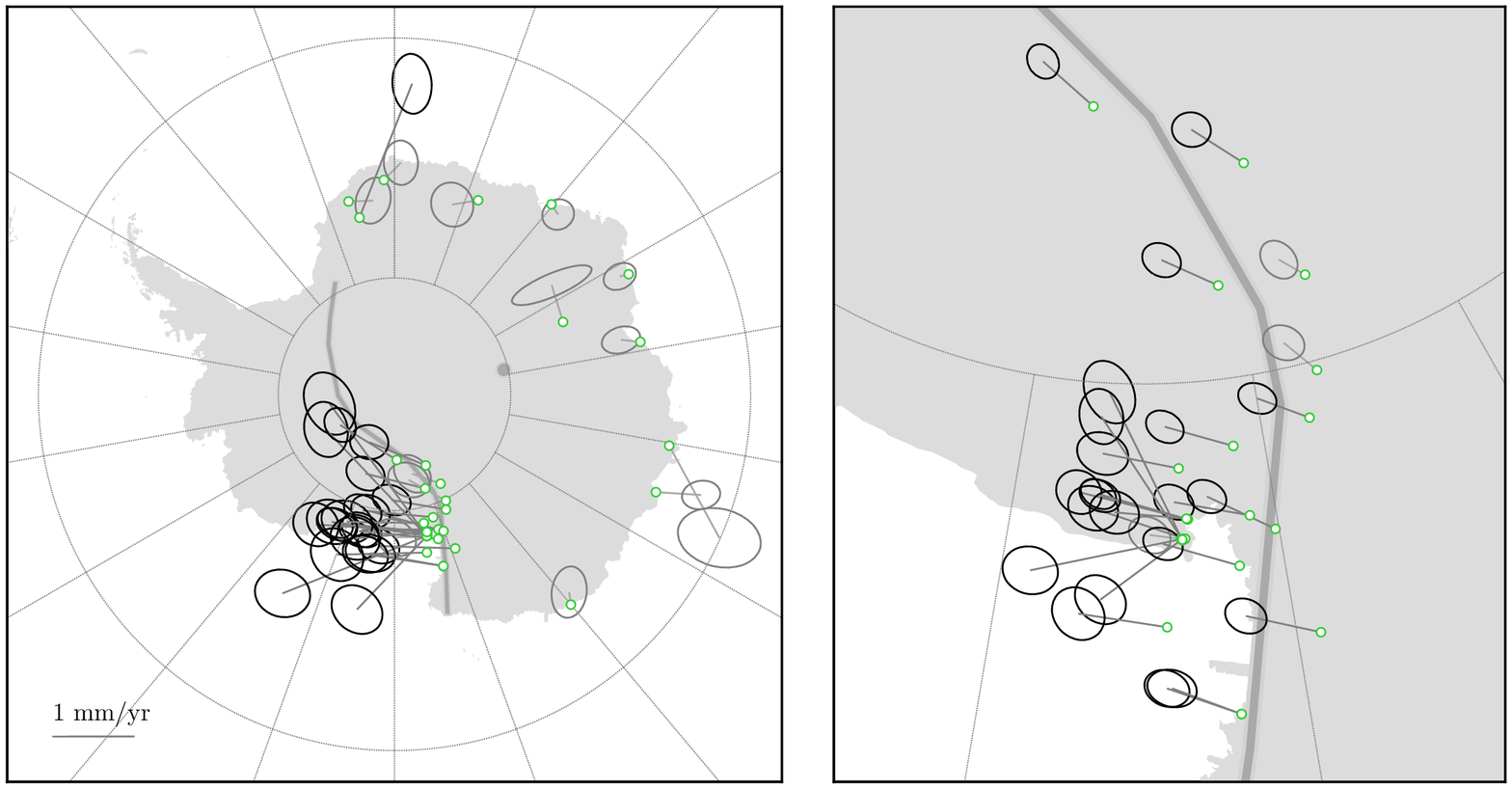}}
\caption{The statistical uncertainties in the {local} velocities derived assuming the weighted regression which corrects for the local number density of stations. The radius of the circles are plotted at the same scale as the vectors and measure the magnitude of the $1\sigma$ uncertainty in the length and direction of the velocity vectors from their best fit values. The vectors and confidence ellipses are shown in black for stations with GIA signals above the measurement and systematic uncertainties at the 2$\sigma$ level; other stations are shown in a lighter grey. A close-up image of the cluster of GPS stations along the Transantarctic Mountains is shown in the right-hand panel with vectors drawn to the same scale as in the left-hand panel.}
\label{fig:Antarctic_GIA_errors}
\end{figure*}

\section{Discussion}
\label{sec:discussion}

\subsection{Station configuration}
\label{sec:Station configuration}

The importance of the distribution and number of GPS stations is explored by repeating the {areal-}weighted regression analysis outlined in Section \ref{sec:simulated} for station configurations used by previous GIA or plate rotation studies. We calculate the median residual between the recovered and simulated `GIA' vector field (simulated using random spherical harmonics) for the various stations distributions on the East Antarctic plate discussed in the literature.
The {areal-}weighted approach not only provides a {lower-noise} estimate of the {local} velocities, but also derives the effective number of independent stations in the separation of the GIA and plate rotation signals. \citet{Altamimi+2002}, \citet{Sella+2002}, \citet{Wei-Ping+2009} include between three and ten East Antarctic GPS stations in their analysis, corresponding to effective counts of less than 7.5 independent stations (Table \ref{tab:distributions}). The station configurations used by \citet{Argus+2014}, and in this work, have effective counts of 10.98 and 11.85 stations respectively.

\begin{table*}
\caption{Theoretical systematic errors (i.e. residuals from 5000 simulated vector fields using random spherical harmonics) in the `GIA' signal arising from various station configurations across the solid Antarctic continental plate. Station distributions are taken from previous plate rotation and GIA studies in the literature. The GPS station in the Kerguelen Islands is isolated from the other East Antarctic sites and may thus be an influence point in the fitting of the plate rotation. The residuals are therefore calculated both with and without this station included in the simulations. These statistics should not be used to compare measurements by these authors; we have not considered factors including the quality of their GPS measurements and whether their sites are located on the rigid plate.}
\small
\centering
\begin{tabular}{l c l c c c}
\hline
 && GPS station count && \multicolumn{2}{c}{GIA velocity residuals (mm/yr)} \\
 && && East Antarctica & EA + Kerguelen \\
\hline
 This work && 36 (or 11.85 effective) && $0.31\pm0.19$ & $0.29\pm0.18$\\
 \citet{Argus+2014} && 23 (or 10.98 effective) && $0.32\pm0.20$ & $0.30\pm0.18$ \\
 \citet{Wei-Ping+2009} && 10 (or 7.42 effective) && $0.37\pm0.22$ & $0.34\pm0.20$ \\
 \citet{Altamimi+2002} && 3 && $0.62\pm0.36$ & $0.49\pm0.30$ \\
% \citet{Larson+1997} && MCM3 && 1 \\
 \citet{Sella+2002} &&  6 (or 5.95 effective) && $0.41\pm0.24$ & $0.36\pm0.23$ \\
 \hline
\end{tabular}
\label{tab:distributions}
\end{table*}

The variation in the velocity residuals between configurations is expected to largely result from the change in this number of independent stations (see Figure \ref{fig:configuration}). The theoretically expected residual velocity at each station for an uncorrelated signal on large length-scales is $\delta v = 1/\sqrt{N}$ mm/yr, where $N$ is the effective number of independent stations. 
The residuals simulated for the station configuration used in this work are consistent with this equation (albeit the median is 6\% higher, {and the rest of distribution similarly shifted}) when using the {areal-}weighted regression. By contrast, the {unweighted} regression has residual velocities 51\% greater than for the weighted regression, and 206\% greater than the theoretical expectation for 36 isolated stations. The median velocity residuals obtained for the station configurations used by \citet{Argus+2014}, \citet{Wei-Ping+2009}, \citet{Altamimi+2002} and \citet{Sella+2002} are also consistent with the expected relationship when using the weighted regression.

\begin{figure}
\centerline{\includegraphics[width=0.9\columnwidth,trim={20 10 40 30},clip]{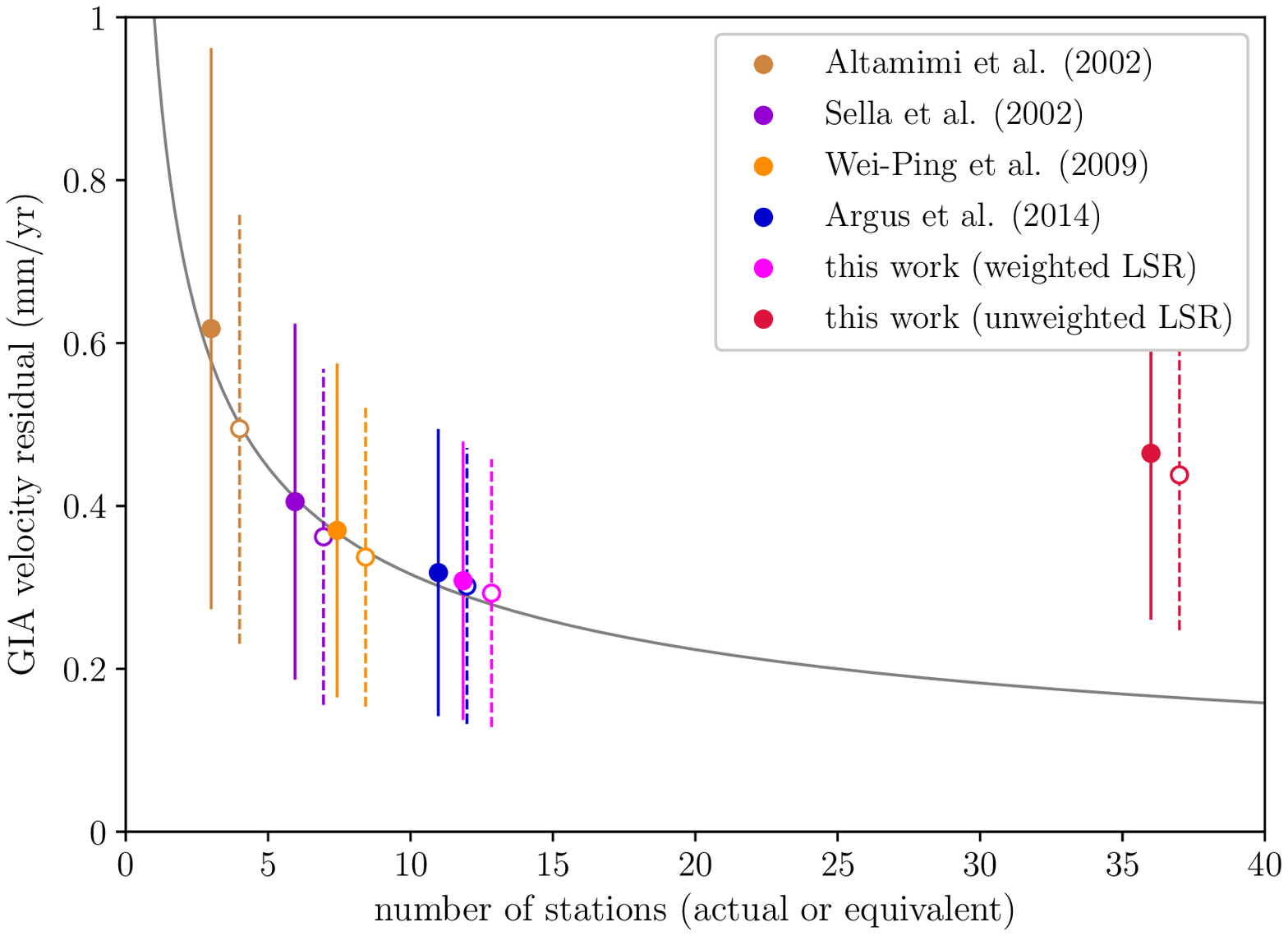}}
\caption{The expected range of (median magnitude) residual velocities simulated for the station configuration in this work and those used by other authors in the literature as a function of the {effective} number of stations. The median and 16/84th percentiles of the distribution of (median magnitude) residual velocities from the 5000 Monte Carlo simulations are shown for each station configuration in East Antarctica ({filled circles/}solid errorbars; colour code described in legend), and also with the inclusion of Kerguelen ({unfilled circles/}dashed errorbars). The theoretical expected residual when fitting a randomly varying signal using an increasing number of stations (solid grey line) is shown for comparison.}
\label{fig:configuration}
\end{figure}

The systematic uncertainty increases greatly with poor spatial sampling due to diminishing confidence in the separation of the tectonic and non-tectonic components of the signal. The simulations undertaken in this work show this uncertainty in the recovered velocity vectors decays from a maximum value (e.g. 1 mm/yr) exponentially with increasing station count. The function reaches a practical limit of diminishing returns (i.e. approaches its asymptotic value) when the total residual falls below a median of 0.2 mm/yr (assuming a median GIA signal of 1 mm/yr), corresponding to an effective station count of 25. For continent-wide studies of East Antarctica, there is therefore an imperative to install 15-20 further, widely spaced, GPS stations to reach this target effective station count. The chosen station configuration of course must also consider practicalities such as the location of bedrock; {this is possible even in the ice-covered Antarctic continent where the are a reasonable number of rocky outcrops with no current GPS instrumentation \citep{Cox+2019}}.  GPS stations in addition to this number would naturally be of benefit to the ongoing study of both vertical and horizontal motions in regions with large variation in their topographical expression and/or likely neotectonic activity.

\subsection{Station velocities from outlying locations}

The Kerguelen Islands, and several other sub-Antarctic islands, have GPS stations sited on the Antarctic Plate but which are potentially less affected by GIA. However, the inclusion of GPS sites at lower latitudes makes the station distribution highly asymmetric and thus their effect on our regression techniques must be carefully considered. 
\citet{Argus+2014} include Kerguelen and Marion Island to obtain their measurement of the GIA signal in Antarctica, whilst most previous authors only include Kerguelen. The Kerguelen Islands formed through intraplate volcanic processes but have had no eruptive volcanic events in recorded history \citep{Ballestracci+1984, Hodgson+2014}. Meanwhile, the glacial history is less well constrained with studies suggesting the main island was completely covered at the Last Glacier Maximum \citep{Hall+1984} whilst others suggest glaciation was limited at this time \citep{Nougier+1972}, potentially supported by the minimal present-day isostatic rebound \citep{Testut+2006}. However, recent rapid deglaciation, possibly linked to increased temperature and decreased precipitation \citep{Frenot+1993}, has decreased the total ice extent on the archipelago from 703 to 552 km$^2$ between 1963 and 2001 \citep{Berthier+2009, Verfaillie+2015}. In contrast, Marion Island is formed by the peaks of an active (eruptions in 1980 and 2004) underwater intraplate shield volcano \citep{McDougall+2001, Hodgson+2014}. Marion Island, and other sub-Antarctic islands, will therefore be not be considered further since their signals will very likely be impacted by local neotectonic or volcanic processes.

We assess the influence outlying stations have on the recovery of the GIA signal by comparing theoretically derived velocity residuals both with and without the inclusion of the Kerguelen GPS station (Table \ref{tab:distributions}). We find the residual velocities are largely unchanged with the inclusion of Kerguelen for the station distribution used in this work, but reduce{d} for the sparser station configurations used by other authors. This improvement is greater than that expected from the increase in station count for the three sparsest station distributions considered (3-8\%), and for the {unweighted} regression applied in this work (5\%; see Figure \ref{fig:configuration}). That is, the inclusion of the distant Kerguelen station provides more leverage in the plate rotation fit than the addition of a site elsewhere on the Antarctic continent. %However, the high leverage of this station is also a disadvantage as any significant non-tectonic motion here will appreciably bias the fit.

The sensitivity of {residual velocities} to the inclusion of Kerguelen is also tested using the measured GPS horizontal velocities. We find the non-plate rotation component of the signal at Kerguelen is $1.14\pm0.51$ mm/yr directed south-east. This relatively large GIA signal (top third of measurements) at an isolated station could bias the fitted plate motion. The angular velocity of the plate rotation increases by 0.01$^\circ/\rm Ma$ and the Euler pole shifts by approximately 1 degree in both latitude and longitude with the inclusion of Kerguelen (Table \ref{tab:plate_rotation2}). The {local} velocities at the other stations are correspondingly subject to moderate changes in magnitude and direction, but the stations identified as having detectable GIA signals remains unchanged.

\subsection{Sites affected by postseismic deformation}

Dumont d'Urville (DUM1) is known to show persistent localised postseismic deformation following the $M_{\rm w}=8.1$ earthquake in 1998, some 600 km northeast of the station along the Macquarie Ridge Complex \citep{King+2016b}. Numerous other large seismic events have occurred along the Macquarie Ridge in the decades prior to 1998, but none have matched the magnitude and proximity to the Antarctic coast \citep{Watson+2010}. There are only episodic GPS measurements at Dumont d'Urville at that epoch, so we instead use observations from DORIS to provide a baseline estimate of the horizontal velocities prior to the earthquake. 

We consider the effect of including sites affected by postseismic deformation (i.e. DUM1) either by using: (1) measurements from before the major seismic event (i.e. using pre-1998 DORIS data, as throughout our results); (2) measurements from after the seismic events (i.e. using 2009-2017 GPS measurements); or (3) removing the station from the analysis. The fitted tectonic plate motions for each of these scenarios is shown in Table \ref{tab:plate_rotation2}. The rotation rate and location of the Euler pole are consistent (differ in the third or fourth significant figure) when using pre-1998 horizontal velocities or excluding the station from the analysis; i.e. the DORIS velocities measured before the seismic event are consistent with the plate motion fitted using the other stations. By contrast, the postseismic horizontal velocities contain a sizable non-tectonic signal, biasing the fitted plate motion significantly. The latitude and longitude differ by 1.1 and 0.4 degrees respectively, though this is within formal errors, whilst the rotation rate is inconsistent at the 2$\sigma$ level. Consideration must therefore be given to which stations may be affected by postseismic deformation, and either removed or use measurements prior to the seismic event.

The postseismic deformation at Dumont d'Urville can be measured directly using the (2009-2017) GPS derived horizontal velocity vector, and the tectonic plate motion derived for the Antarctic continent using the 35 other stations. The postseismic velocity vector at Dumont d'Urville is $\boldsymbol{v_{\rm tot}} = -12.36\boldsymbol{e_\theta} + 7.52\boldsymbol{e_\phi}$ mm/yr whilst the fitted component due to plate rotation is $\boldsymbol{v} = -12.03\boldsymbol{e_\theta} + 6.12\boldsymbol{e_\phi}$ mm/yr. Postseismic deformation, and the much weaker glacial rebound, thus leads to horizontal motion at Dumont d'Urville of $1.43\pm0.75$ mm/yr directed at a bearing of 103 degrees. {Eastward motion is plausibly explained by the strike-slip mechanism on an east-west fault plane \citep{Nettles+1999, King+2016b}.}

\begin{table*}
\caption{Rotation rates and Euler poles calculated {from {observational} data} for the Antarctic Plate using the weighted LSR including: (1) all East Antarctic GPS stations but using DORIS at DUM1 (best case); (2) all East Antarctic stations except for DUM1; (3) all East Antarctic stations with `incorrect' post-seismic GPS data at DUM1; or (4) both Kerguelen and the East Antarctic stations. {The Euler pole parameters of \citet{Argus+2014} and \citet[][both derived using KERG]{Wei-Ping+2009} have also been provided for comparision with our fit including Kerguelen.} The columns are the same as for Table \ref{tab:plate_rotation}.}
\small
\centering
\begin{tabular}{l c c c c c c}
\hline
 & Latitude & Longitude & \multicolumn{4}{c}{Angular velocity}  \\
 & $\theta_\epsilon$ ($^\circ$N) & $\phi_\epsilon$ ($^\circ$E) & $\omega_{\rm x}$ (mas/a) & $\omega_{\rm y}$ (mas/a) & $\omega_{\rm z}$ (mas/a) & $\omega$ ($^\circ/\rm Ma$) \\
\hline
  East Antarctica  & $55.04_{-0.72}^{+0.65}$ & $-126.17_{-0.91}^{+0.93}$ & \!\!$-0.232\pm0.007$\! & \!\!$-0.318\pm0.008$\! & $0.563\pm0.022$ & $0.191\pm0.006$ \\
  EA (excl. DUM1) & $55.07_{-0.66}^{+0.64}$ & $-126.24_{-0.93}^{+0.93}$ & \!\!$-0.231\pm0.008$\! & \!\!$-0.316\pm0.008$\! & $0.560\pm0.024$ & $0.190\pm0.007$ \\
  EA (GPS at DUM1) & $56.15_{-0.57}^{+0.55}$ & $-126.61_{-0.89}^{+0.86}$ & \!\!$-0.232\pm0.007$\! & \!\!$-0.318\pm0.007$\! & $0.563\pm0.022$ & $0.201\pm0.006$ \\
  EA + Kerguelen & $56.60_{-0.44}^{+0.42}$ & $-126.26_{-0.82}^{+0.81}$ & \!\!$-0.238\pm0.006$\! & \!\!$-0.324\pm0.007$\! & $0.609\pm0.017$ & $0.203\pm0.005$ \\
\hline
  \citet{Argus+2014}  & $59.876$ & $-127.277$ & -- & -- & -- & $0.2178$ \\
  \citet{Wei-Ping+2009}  & $58.69\pm0.4$ & $-128.29\pm0.3$ & -- & -- & -- & $0.224\pm0.01$ \\
\hline
\end{tabular}
\label{tab:plate_rotation2}
\end{table*}

\section{Conclusion}
\label{sec:conclusion}

{We have demonstrated that an {areal-}weighted regression, using an increased number of stations, is a well-posed approach for the separation of plate rotation and more local components of horizontal GPS stations velocities, such as that due to glacial isostatic adjustment. Our approach takes into account likely correlation length-scale and quantifies the systematic errors in fitting plate rotation in the presence of a strong GIA signal. A careful appraisal of error propagation using a Monte Carlo approach allows the identification of stations with local horizontal components that may be regarded as significant.  As a case example, we provide a list of {local} velocity vectors for GPS station sites in East Antarctica, noting that 25 of which are significant at the 2$\sigma$ confidence level for the data available at the time of calculation. These can be used in ongoing GIA studies to inform models of the local Earth rheology and deglaciation history.}

{The techniques discussed in this work can be applied in other regions globally (e.g. Greenland, Europe, North America) to constrain the signal from glacial rebound. We have made the code used in our theoretical simulations publically available in the supplementary material. This code provides estimates of the residual and recovered signal for a user defined station configuration sited on one of the major tectonic plates (and for any specified Euler pole and rotation rate). The code may potentially be used in the optimisation of future additions to regional or continental scale GPS deployments.}

\section*{Acknowledgments}
This research was supported under Australian Research Council's Special Research Initiative for Antarctic Gateway Partnership (Project ID SR140300001), Discovery Project DP170100224 and by Department of Environment and Energy Australian Antarctic Science Program grant 4318. We thank Pippa Whitehouse for providing GIA model output and NASA JPL for making GIPSY and associated orbit products available. {We further thank Duncan Agnew, Jeffrey Freymueller and an anonymous reviewer for their helpful and constructive feedback that has improved this manuscript.}

\bibliographystyle{apalike}
\bibliography{signalseparation}

\begin{thebibliography}{}

\bibitem[Altamimi et~al., 2011]{Altamimi+2011}
Altamimi, Z., Collilieux, X., and M{\'e}tivier, L. (2011).
\newblock {ITRF2008: an improved solution of the international terrestrial
  reference frame}.
\newblock {\em Journal of Geodesy}, 85(8):457--473.

\bibitem[Altamimi et~al., 2012]{Altamimi+2012}
Altamimi, Z., M{\'e}tivier, L., and Collilieux, X. (2012).
\newblock {ITRF2008 plate motion model}.
\newblock {\em Journal of Geophysical Research: Solid Earth}, 117(B7).

\bibitem[Altamimi et~al., 2017]{Altamimi+2017}
Altamimi, Z., M{\'e}tivier, L., Rebischung, P., Rouby, H., and Collilieux, X.
  (2017).
\newblock {ITRF2014 plate motion model}.
\newblock {\em Geophysical Journal International}, 209(3):1906--1912.

\bibitem[Altamimi et~al., 2002]{Altamimi+2002}
Altamimi, Z., Sillard, P., and Boucher, C. (2002).
\newblock {ITRF2000: A new release of the International Terrestrial Reference
  Frame for earth science applications}.
\newblock {\em Journal of Geophysical Research: Solid Earth}, 107(B10):ETG
  2--1--ETG 2--19.

\bibitem[Argus et~al., 2014]{Argus+2014}
Argus, D.~F., Peltier, W.~R., Drummond, R., and Moore, A.~W. (2014).
\newblock {The Antarctica component of postglacial rebound model ICE-6G\_C
  (VM5a) based on GPS positioning, exposure age dating of ice thicknesses, and
  relative sea level histories}.
\newblock {\em Geophysical Journal International}, 198(1):537--563.

\bibitem[Ballestracci and Nougier, 1984]{Ballestracci+1984}
Ballestracci, R. and Nougier, J. (1984).
\newblock {Detection by infrared thermography and modelling of an icecapped
  geothermal system in Kerguelen archipelago}.
\newblock {\em Journal of Volcanology and Geothermal Research}, 20(1):85 -- 99.

\bibitem[Barletta et~al., 2018]{Barletta+2018}
Barletta, V.~R., Bevis, M., Smith, B.~E., Wilson, T., Brown, A., Bordoni, A.,
  Willis, M., Khan, S.~A., Rovira-Navarro, M., Dalziel, I., Smalley, R.,
  Kendrick, E., Konfal, S., Caccamise, D.~J., Aster, R.~C., Nyblade, A., and
  Wiens, D.~A. (2018).
\newblock {Observed rapid bedrock uplift in Amundsen Sea Embayment promotes
  ice-sheet stability}.
\newblock {\em Science}, 360(6395):1335--1339.

\bibitem[Barletta et~al., 2006]{Barletta+2006}
Barletta, V.~R., Ferrari, C., Diolaiuti, G., Carnielli, T., Sabadini, R., and
  Smiraglia, C. (2006).
\newblock {Glacier shrinkage and modeled uplift of the Alps}.
\newblock {\em Geophysical Research Letters}, 33(14).

\bibitem[Berthier et~al., 2009]{Berthier+2009}
Berthier, E., Le~Bris, R., Mabileau, L., Testut, L., and R\'{e}my, F. (2009).
\newblock {Ice wastage on the Kerguelen Islands (49$^\circ$S, 699$^\circ$E)
  between 1963 and 2006}.
\newblock {\em Journal of Geophysical Research: Earth Surface}, 114(F3).

\bibitem[Cox et~al., 2019]{Cox+2019}
Cox, S.~C., Smith, L.~B., and {the GeoMAP team} (2019).
\newblock {Geological dataset of Antarctica, GeoMAP.v.201907}.

\bibitem[DeMets et~al., 2016]{DeMets+2016}
DeMets, C., Calais, E., and Merkouriev, S. (2016).
\newblock {Reconciling geodetic and geological estimates of recent plate motion
  across the Southwest Indian Ridge}.
\newblock {\em Geophysical Journal International}, 208(1):118--133.

\bibitem[Frenot et~al., 1993]{Frenot+1993}
Frenot, Y., Gloaguen, J., Picot, G., Bougère, J., and Benjamin, D. (1993).
\newblock {Azorella selago Hook. used to estimate glacier fluctuations and
  climatic history in the Kerguelen Islands over the last two centuries}.
\newblock {\em Oecologia}, 95(1):140--144.

\bibitem[Fretwell et~al., 2013]{Fretwell+2013}
Fretwell, P., Pritchard, H.~D., Vaughan, D.~G., Bamber, J.~L., Barrand, N.~E.,
  Bell, R., Bianchi, C., Bingham, R.~G., Blankenship, D.~D., Casassa, G.,
  Catania, G., Callens, D., Conway, H., Cook, A.~J., Corr, H. F.~J., Damaske,
  D., Damm, V., Ferraccioli, F., Forsberg, R., Fujita, S., Gim, Y., Gogineni,
  P., Griggs, J.~A., Hindmarsh, R. C.~A., Holmlund, P., Holt, J.~W., Jacobel,
  R.~W., Jenkins, A., Jokat, W., Jordan, T., King, E.~C., Kohler, J., Krabill,
  W., Riger-Kusk, M., Langley, K.~A., Leitchenkov, G., Leuschen, C., Luyendyk,
  B.~P., Matsuoka, K., Mouginot, J., Nitsche, F.~O., Nogi, Y., Nost, O.~A.,
  Popov, S.~V., Rignot, E., Rippin, D.~M., Rivera, A., Roberts, J., Ross, N.,
  Siegert, M.~J., Smith, A.~M., Steinhage, D., Studinger, M., Sun, B., Tinto,
  B.~K., Welch, B.~C., Wilson, D., Young, D.~A., Xiangbin, C., and Zirizzotti,
  A. (2013).
\newblock {Bedmap2: improved ice bed, surface and thickness datasets for
  Antarctica}.
\newblock {\em The Cryosphere}, 7(1):375--393.

\bibitem[Ghavri et~al., 2015]{Ghavri+2015}
Ghavri, S., Catherine, J.~K., and Gahalaut, V.~K. (2015).
\newblock {First estimate of plate motion at Maitri GPS site, Indian base
  station at Antarctica}.
\newblock {\em Journal of the Geological Society of India}, 85(4):431--433.

\bibitem[Goudarzi et~al., 2014]{Goudarzi+2014}
Goudarzi, M., Cocard, M., and Santerre, R. (2014).
\newblock {EPC: Matlab software to estimate Euler pole parameters}.
\newblock {\em GPS Solutions}, 18.

\bibitem[Hall and Visser, 1984]{Hall+1984}
Hall, K. and Visser, J. (1984).
\newblock {Observations on the relationship between clast size, shape, and
  lithology from the Permo-Carboniferous glaciogenic Dwyka Formation in the
  western part of the Karoo Basin}.
\newblock {\em Transactions - Geological Society of South Africa},
  87(3):225--232.

\bibitem[Harley et~al., 2013]{Harley+2013}
Harley, S.~L., Fitzsimons, I. C.~W., and Zhao, Y. (2013).
\newblock {Antarctica and supercontinent evolution: historical perspectives,
  recent advances and unresolved issues}.
\newblock {\em Geological Society, London, Special Publications}, 383(1):1--34.

\bibitem[Hirth and Kohlstedt, 2013]{Hirth+2013}
Hirth, G. and Kohlstedt, D. (2013).
\newblock {\em {Rheology of the Upper Mantle and the Mantle Wedge: A View from
  the Experimentalists}}, pages 83--105.
\newblock American Geophysical Union (AGU).

\bibitem[Hodgson et~al., 2014]{Hodgson+2014}
Hodgson, D.~A., Graham, A.~G., Roberts, S.~J., Bentley, M.~J., Cofaigh, C.~O.,
  Verleyen, E., Vyverman, W., Jomelli, V., Favier, V., Brunstein, D.,
  Verfaillie, D., Colhoun, E.~A., Saunders, K.~M., Selkirk, P.~M., Mackintosh,
  A., Hedding, D.~W., Nel, W., Hall, K., McGlone, M.~S., der Putten, N.~V.,
  Dickens, W.~A., and Smith, J.~A. (2014).
\newblock {Terrestrial and submarine evidence for the extent and timing of the
  Last Glacial Maximum and the onset of deglaciation on the maritime-Antarctic
  and sub-Antarctic islands}.
\newblock {\em Quaternary Science Reviews}, 100:137 -- 158.

\bibitem[Johansson et~al., 2002]{Johansson+2002}
Johansson, J.~M., Davis, J.~L., Scherneck, H.-G., Milne, G.~A., Vermeer, M.,
  Mitrovica, J.~X., Bennett, R.~A., Jonsson, B., Elgered, G., Elósegui, P.,
  Koivula, H., Poutanen, M., Rönnäng, B.~O., and Shapiro, I.~I. (2002).
\newblock {Continuous GPS measurements of postglacial adjustment in
  Fennoscandia 1. Geodetic results}.
\newblock {\em Journal of Geophysical Research: Solid Earth}, 107(B8):ETG
  3--1--ETG 3--27.

\bibitem[Jordan et~al., 2020]{Jordan+2020}
Jordan, T.~A., Riley, T.~R., and Siddoway, C.~S. (2020).
\newblock {The geological history and evolution of West Antarctica}.
\newblock {\em Nature Reviews Earth \& Environment}, 1(2):117--133.

\bibitem[King et~al., 2012]{King+2012}
King, M.~A., Bingham, R.~J., Moore, P., Whitehouse, P.~L., Bentley, M.~J., and
  Milne, G.~A. (2012).
\newblock {Lower satellite-gravimetry estimates of Antarctic sea-level
  contribution}.
\newblock {\em Nature}, 491(7425):586.

\bibitem[King and Santamar\'{i}a-G\'{o}mez, 2016]{King+2016b}
King, M.~A. and Santamar\'{i}a-G\'{o}mez, A. (2016).
\newblock {Ongoing deformation of Antarctica following recent Great
  Earthquakes}.
\newblock {\em Geophysical Research Letters}, 43(5):1918--1927.

\bibitem[King et~al., 2016]{King+2016}
King, M.~A., Whitehouse, P.~L., and van~der Wal, W. (2016).
\newblock {Incomplete separability of Antarctic plate rotation from glacial
  isostatic adjustment deformation within geodetic observations}.
\newblock {\em Geophysical Journal International}, 204(1):324--330.

\bibitem[Klemann et~al., 2008]{Klemann+2008}
Klemann, V., Martinec, Z., and Ivins, E.~R. (2008).
\newblock {Glacial isostasy and plate motion}.
\newblock {\em Journal of Geodynamics}, 46(3):95 -- 103.
\newblock Glacial Isostatic Adjustment: New Developments and Applications in
  Global Change, Hydrology, Sea Level, Cryosphere, and Geodynamics.

\bibitem[Kreemer et~al., 2018]{Kreemer+2018}
Kreemer, C., Hammond, W.~C., and Blewitt, G. (2018).
\newblock {A Robust Estimation of the 3-D Intraplate Deformation of the North
  American Plate From GPS}.
\newblock {\em Journal of Geophysical Research: Solid Earth},
  123(5):4388--4412.

\bibitem[Larson et~al., 1997]{Larson+1997}
Larson, K.~M., Freymueller, J.~T., and Philipsen, S. (1997).
\newblock {Global plate velocities from the Global Positioning System}.
\newblock {\em Journal of Geophysical Research: Solid Earth},
  102(B5):9961--9981.

\bibitem[Latychev et~al., 2005]{Latychev+2005}
Latychev, K., Mitrovica, J.~X., Tamisiea, M.~E., Tromp, J., and Moucha, R.
  (2005).
\newblock {Influence of lithospheric thickness variations on 3-D crustal
  velocities due to glacial isostatic adjustment}.
\newblock {\em Geophysical Research Letters}, 32(1).

\bibitem[McDougall et~al., 2001]{McDougall+2001}
McDougall, I., Verwoerd, W., and Chevallier, L. (2001).
\newblock {K-Ar geochoronology of Marion Island, Southern Ocean}.
\newblock {\em Geological Magazine}, 138(1):1--17.

\bibitem[McKenzie and Parker, 1974]{McKenzie+1974}
McKenzie, D. and Parker, R. (1974).
\newblock {Plate tectonics in $\omega$ space}.
\newblock {\em Earth and Planetary Science Letters}, 22(4):285--293.

\bibitem[Nettles et~al., 1999]{Nettles+1999}
Nettles, M., Wallace, T.~C., and Beck, S.~L. (1999).
\newblock {The March 25, 1998 Antarctic Plate Earthquake}.
\newblock {\em Geophysical Research Letters}, 26(14):2097--2100.

\bibitem[Nield et~al., 2014]{Nield+2014}
Nield, G.~A., Barletta, V.~R., Bordoni, A., King, M.~A., Whitehouse, P.~L.,
  Clarke, P.~J., Domack, E., Scambos, T.~A., and Berthier, E. (2014).
\newblock {Rapid bedrock uplift in the Antarctic Peninsula explained by
  viscoelastic response to recent ice unloading}.
\newblock {\em Earth and Planetary Science Letters}, 397:32 -- 41.

\bibitem[Nougier, 1972]{Nougier+1972}
Nougier, J. (1972).
\newblock {Aspects de morpho-tectonique glaciaire aux Iles Kerguelen}.
\newblock {\em Revue de G\'{e}ographie Physique et de G\'{e}ologie Dynamique},
  14:499--505.

\bibitem[Peltier et~al., 2015]{Peltier+2015}
Peltier, W.~R., Argus, D.~F., and Drummond, R. (2015).
\newblock {Space geodesy constrains ice age terminal deglaciation: The global
  ICE-6G\_C (VM5a) model}.
\newblock {\em Journal of Geophysical Research: Solid Earth}, 120(1):450--487.

\bibitem[Ritsema et~al., 2011]{Ritsema+2011}
Ritsema, J., Deuss, A., van Heijst, H.~J., and Woodhouse, J.~H. (2011).
\newblock {S40RTS: a degree-40 shear-velocity model for the mantle from new
  Rayleigh wave dispersion, teleseismic traveltime and normal-mode splitting
  function measurements}.
\newblock {\em Geophysical Journal International}, 184(3):1223--1236.

\bibitem[Roberts et~al., 2018]{Roberts+2018}
Roberts, J., Galton-Fenzi, B.~K., Paolo, F.~S., Donnelly, C., Gwyther, D.~E.,
  Padman, L., Young, D., Warner, R., Greenbaum, J., Fricker, H.~A., Payne,
  A.~J., Cornford, S., Le~Brocq, A., van Ommen, T., Blankenship, D., and
  Siegert, M.~J. (2018).
\newblock {Ocean forced variability of Totten Glacier mass loss}.
\newblock {\em Geological Society, London, Special Publications},
  461(1):175--186.

\bibitem[Sella et~al., 2002]{Sella+2002}
Sella, G.~F., Dixon, T.~H., and Mao, A. (2002).
\newblock {REVEL: A model for Recent plate velocities from space geodesy}.
\newblock {\em Journal of Geophysical Research: Solid Earth}, 107(B4):ETG
  11--1--ETG 11--30.

\bibitem[Sella et~al., 2007]{Sella+2007}
Sella, G.~F., Stein, S., Dixon, T.~H., Craymer, M., James, T.~S., Mazzotti, S.,
  and Dokka, R.~K. (2007).
\newblock {Observation of glacial isostatic adjustment in “stable” North
  America with GPS}.
\newblock {\em Geophysical Research Letters}, 34(2).

\bibitem[Serpelloni et~al., 2013]{Serpelloni+2013}
Serpelloni, E., Faccenna, C., Spada, G., Dong, D., and Williams, S. D.~P.
  (2013).
\newblock {Vertical GPS ground motion rates in the Euro-Mediterranean region:
  New evidence of velocity gradients at different spatial scales along the
  Nubia-Eurasia plate boundary}.
\newblock {\em Journal of Geophysical Research: Solid Earth},
  118(11):6003--6024.

\bibitem[Siddoway, 2008]{Siddoway+2008}
Siddoway, C.~S. (2008).
\newblock {Tectonics of the West Antarctic Rift System: new light on the
  history and dynamics of distributed intracontinental extension}.
\newblock {\em Antarctica: A Keystone in a Changing World}, pages 91--114.

\bibitem[Testut et~al., 2006]{Testut+2006}
Testut, L., W{\"o}ppelmann, G., Simon, B., and T{\'e}chin{\'e}, P. (2006).
\newblock {The sea level at Port-aux-Fran{\c{c}}ais, Kerguelen Island, from
  1949 to the present}.
\newblock {\em Ocean Dynamics}, 56(5):464--472.

\bibitem[van~der Wal et~al., 2015]{vanderWal+2015}
van~der Wal, W., Whitehouse, P.~L., and Schrama, E.~J. (2015).
\newblock {Effect of GIA models with 3D composite mantle viscosity on GRACE
  mass balance estimates for Antarctica}.
\newblock {\em Earth and Planetary Science Letters}, 414:134 -- 143.

\bibitem[van~der Wal et~al., 2010]{vanderWal+2010}
van~der Wal, W., Wu, P., Wang, H., and Sideris, M.~G. (2010).
\newblock {Sea levels and uplift rate from composite rheology in glacial
  isostatic adjustment modeling}.
\newblock {\em Journal of Geodynamics}, 50(1):38 -- 48.

\bibitem[Verfaillie et~al., 2015]{Verfaillie+2015}
Verfaillie, D., Favier, V., Dumont, M., Jomelli, V., Gilbert, A., Brunstein,
  D., Gallée, H., Rinterknecht, V., Menegoz, M., and Frenot, Y. (2015).
\newblock {Recent glacier decline in the Kerguelen Islands (49$\circ$S,
  69$\circ$E) derived from modeling, field observations, and satellite data}.
\newblock {\em Journal of Geophysical Research: Earth Surface},
  120(3):637--654.

\bibitem[Watson et~al., 2010]{Watson+2010}
Watson, C., Burgette, R., Tregoning, P., White, N., Hunter, J., Coleman, R.,
  Handsworth, R., and Brolsma, H. (2010).
\newblock {Twentieth century constraints on sea level change and earthquake
  deformation at Macquarie Island}.
\newblock {\em Geophysical Journal International}, 182(2):781--796.

\bibitem[Wei-Ping et~al., 2009]{Wei-Ping+2009}
Wei-Ping, J.~E., Dong-Chen, Z. B.-., and Liu, Y. (2009).
\newblock {New Model of Antarctic Plate Motion and Its Analysis}.
\newblock {\em Chinese Journal of Geophysics}, 52(1):23--32.

\bibitem[Whitehouse et~al., 2012]{Whitehouse+2012}
Whitehouse, P.~L., Bentley, M.~J., Milne, G.~A., King, M.~A., and Thomas, I.~D.
  (2012).
\newblock {A new glacial isostatic adjustment model for Antarctica: calibrated
  and tested using observations of relative sea-level change and present-day
  uplift rates}.
\newblock {\em Geophysical Journal International}, 190(3):1464--1482.

\bibitem[Williams, 2008]{Williams+2008}
Williams, S. D.~P. (2008).
\newblock {CATS: GPS coordinate time series analysis software}.
\newblock {\em GPS Solutions}, 12(2):147--153.

\bibitem[Wolstencroft et~al., 2015]{Wolstencroft+2015}
Wolstencroft, M., King, M.~A., Whitehouse, P.~L., Bentley, M.~J., Nield, G.~A.,
  King, E.~C., McMillan, M., Shepherd, A., Barletta, V., Bordoni, A., Riva,
  R.~E., Didova, O., and Gunter, B.~C. (2015).
\newblock {Uplift rates from a new high-density GPS network in Palmer Land
  indicate significant late Holocene ice loss in the southwestern Weddell Sea}.
\newblock {\em Geophysical Journal International}, 203(1):737--754.

\end{thebibliography}

%\bsp % ``This paper has been produced using the Blackwell
     %   Publishing GJI \LaTeXe\ class file.''

\label{lastpage}

\end{document}